\documentclass[eqsecnum,fleqn,twoside,twocolumn,nofootinbib,showkeys]{revtex4}
\usepackage[nocpr]{ujp}

\usepackage{dsfont}

\newcommand\I{\texttt{i}}
\newcommand\D{\mathrm{d}}
\newcommand\E{\mathrm{e}}
\renewcommand\O{\mathcal{O}}
\renewcommand\Re{\mathop{\rm Re}\nolimits}
\newcommand\refe[1]{(\ref{#1})}

\begin{document}
\title[High-Order Nonlinear Schr\"{o}dinger Equation]
{HIGH-ORDER NONLINEAR\\ SCHR\"{O}DINGER EQUATION FOR THE ENVELOPE\\ OF SLOWLY MODULATED GRAVITY WAVES\\ ON THE SURFACE OF FINITE-DEPTH FLUID\\ AND ITS QUASI-SOLITON SOLUTIONS}
\author{I.S.~Gandzha}
\affiliation{\iop}
\address{\iopaddr}
\email{gandzha@iop.kiev.ua, sedlets@iop.kiev.ua}
\author{Yu.V.~Sedletsky}
\affiliation{\iop}
\address{\iopaddr}
\author{D.S.~Dutykh\,}
\affiliation{Universit\'e de Savoie Mont Blanc}
\address{CNRS--LAMA UMR 5127, Campus Universitaire, 73376 Le Bourget-du-Lac,\\ \hspace*{1mm} France}
\email{Denys.Dutykh@univ-savoie.fr}

\udk{532.591} \razd{\secx}
\autorcol{I.S.\hspace*{0.7mm}Gandzha, Yu.V.\hspace*{0.7mm}Sedletsky, D.S.\hspace*{0.7mm}Dutykh}%

\setcounter{page}{1201}


\begin{abstract}
We consider the high-order nonlinear Schr\"{o}dinger equation
derived earlier by Sedletsky [Ukr.\,\,J.\,\,Phys.\,\,\textbf{48}(1),
82 (2003)] for the first-harmonic envelope of slowly modulated
gravity waves on the surface of finite-depth irrotational, inviscid,
and incompressible fluid with flat bottom.\,\,This equation takes
into account the third-order dispersion and cubic nonlinear
dispersive terms.\,\,We rewrite this equation in dimensionless form
featuring only one dimensionless parameter $kh$, where $k$ is the
carrier wavenumber and $h$ is the undisturbed fluid depth.\,\,We
show that one-soliton solutions of the classical nonlinear
Schr\"{o}dinger equation are transformed into quasi-soliton
solutions with slowly varying amplitude when the high-order terms
are taken into consideration.\,\,These quasi-soliton solutions
represent the secondary modulations of gravity waves.
\end{abstract}
\maketitle

\section{Introduction}
The nonlinear Schr\"{o}dinger equation (NLSE)
\begin{equation}\label{eq:NLSE}
A_\tau = - a_1 A_\chi -\I a_2 A_{\chi\chi} +\I a_{0,\,0,\,0}A|A|^2
\end{equation}
arises in describing nonlinear waves in various physical contexts,
such as nonlinear optics \cite{Turitsyn_Review_2012}, plasma physics
\cite{Infeld_Rowlands_2000}, nanosized electronics
\cite{Crutcher_2012}, ferromagnetics \cite{chen1993}, Bose--Einstein
condensates \cite{Zakharov_Review_2012}, and hydrodynamics
\cite{Zakharov_1971, Yuen_Lake_1975, Dias_Kharif_Review_1999,
Onorato_Review_2013}.\,\,Here, $\chi$ is the direction of wave
propagation, $\tau$ is time, $A(\chi,\,\tau)$ is the complex
first-harmonic envelope of the carrier wave, and the subscripts next
to $A$ denote the partial derivatives.\,\,NLSE takes into account
the second-order dispersion (term with $A_{\chi\chi}$) and the phase
self-modulation (term with $A|A|^2$).\,\,The coefficients $a_1$,
$a_2$, and $a_{0,\,0,\,0}$ take various values depending on the
particular physical context under consideration.

In the general context of weakly nonlinear dispersive waves, this
equation was first discussed by Benney and Newell
\cite{Benney_Newell_1967}.\,\,In the case of gravity waves
propagating on the surface of infinite-depth irrotational, inviscid,
and incompressible fluid, NLSE was first derived by Zakharov
\cite{Zakharov_1968} using the Hamiltonian formalism and then by
Yuen and Lake \cite{Yuen_Lake_1975} using the averaged Lagrangian
method.\,\,The finite-depth NLSE of form \refe{eq:NLSE} was first
derived by Hasimoto and Ono \cite{Hasimoto_Ono_1972} using the
multiple-scale method and then by Stiassnie and Shemer
\cite{Stiassnie_Shemer_1984} from Zakharov's integral equations.
Noteworthy is also the recent paper by Thomas et al.
\cite{Thomas_Kharif_2012} who derived the finite-depth NLSE for
water waves on finite depth with constant vorticity.

Under certain relationship between the parameters,
when\vspace*{-2mm}
\begin{equation}\label{eq:MI}
 a_2 a_{0,\,0,\,0}\ <\ 0,
\end{equation}
NLSE admits exact solutions in the form of solitons, which exist due
to the balance of dispersion and nonlinearity and propagate without
changing their shape and  keeping their energy
\cite{Dodd_solitons}.\,\,In this case, the uniform carrier wave is
unstable with respect to long-wave modulations allowing for the
formation of envelope solitons.\,\,This type of instability is known
as the modulational or Benjamin--Feir instability
\cite{Zakharov_MI_2009} (it was discovered for the first time in
optics by Bespalov and Talanov \cite{Bespalov_1966}).\,\,In the case
of surface gravity waves, condition \refe{eq:MI} holds at
$kh\gtrsim1.363$, $k$ being the carrier wavenumber and $h$ being the
undisturbed fluid depth.\,\,In addition to theoretical predictions,
envelope solitons were observed in numerous experiments performed in
water tanks
\cite{Yuen_Lake_1975,Yuen_Lake_1982,Su_1982,Shemer_2010,Chabchoub_2011,Onorato_Review_2013,
Slunyaev_PF_2013,Shemer_2013}.

At the bifurcation point $a_{0,\,0,\,0} = 0$ ($kh\approx 1.363$),
when the modulational instability changes to stability, NLSE of form
\refe{eq:NLSE} is not sufficient to describe the wavetrain evolution
since the leading nonlinear term vanishes.\,\,In this case,
high-order nonlinear and nonlinear-dispersive terms should be taken
into account.\,\,In the case of infinite depth, such a high-order
NLSE (HONLSE) was first derived by Dysthe \cite{Dysthe_1979}.\,\,It
includes the third-order dispersion ($A_{\chi\chi\chi}$) and cubic
nonlinear dispersive terms ($|A|^2 A_{\chi}$, $A^2 A^*_{\chi}$,
asterisk denotes the complex conjugate) as well as an additional
nonlinear dispersive term describing the input of the wave-induced
mean flow (some of these terms were introduced earlier by Roskes
\cite{Roskes_1977} without taking into consideration the induced
mean flow).\,\,This equation is usually referred to as the
fourth-order HONLSE to emphasize the contrast with the third-order
NLSE.\,\,Janssen \cite{Janssen_1983} re-derived Dysthe's equation
and corrected the sign at one of the nonlinear dispersive terms.
Hogan \cite{Hogan_1985} followed the earlier work by Stiassnie
\cite{Stiassnie_1984} to derive the similar equation for deep-water
gravity-capillary waves with surface tension taken into
account.\,\,Selezov et al.\,\,\cite{Selezov_2003} extended the
HONLSE derived by Hogan to the case of nonlinear wavetrain
propagation on the interface of two semi-infinite fluids without
taking into account the induced mean flow.\,\,Worthy of mention is
also the paper by Lukomsky \cite{Lukom_1995} who derived Dysthe's
equation in a different way.\,\,Later, Trulsen and Dysthe
\cite{Trulsen_Dysthe_1996} extended the equation derived by Dysthe
to broader bandwidth by including the fourth- and fifth-order linear
dispersion.\,\,Debsarma and Das \cite{Debsarma_Das_2005} derived a
yet more general HONLSE that is one order higher than the equation
derived by Trulsen and Dysthe.\,\,Gramstad and Trulsen
\cite{Gramstad_Trulsen_PF_2011} derived a set of two coupled
fourth-order HONLSEs capable of describing two interacting wave
systems separated in wavelengths or directions of
propagation.\,\,Zakharov and Dyachenko
\cite{Zakharov_Dyachenko_2010,Zakharov_Dyachenko_2011,Zakharov_Dyachenko_2012}
made a conformal mapping of the fluid domain to the lower half-plane
to derive a counterpart of Dysthe's equation in new canonical
variables (the so-called compact Dyachenko--Zakharov equation
\mbox{\cite{Fedele_Dutykh_2012_JETP,Fedele_Dutykh_2012}).}\looseness=1

Original Dysthe's equation was written for the first-harmonic
envelope of velocity potential rather than of surface profile.\,\,In
the case of standard NLSE, this difference in not essential because
in that order the first-harmonic amplitudes of the velocity
potential and surface displacement differ by a dimensional factor
only, which is not true anymore in the HONLSE case, as discussed by
Hogan \cite{Hogan_1986}.\,\,Keeping this in mind, Trulsen et al.
\cite{Trulsen_Dysthe_2000} rewrote Dysthe's equation in terms of the
first-harmonic envelope of surface profile while taking into account
the linear dispersion to an arbitrary order.

In the case of finite depth, the effect of induced mean flow
manifests itself in the third order, so that the NLSE is generally
coupled to the equation for the induced mean flow
\cite{Benney_Roskes_1969}.\,\,However, Davey and Stewartson
\cite{Davey_Stewartson_1974} showed that these coupled equations are
equivalent to the single NLSE derived by Hasimoto and Ono
\cite{Hasimoto_Ono_1972}.\,\,On the other hand, such an equivalence
is not preserved for high-order equations.\,\,The first attempt to
derive a HONLSE in the case of finite depth was made by Johnson
\cite{Johnson_1977}, but only for $kh\approx 1.363$, when the cubic
NLSE term vanishes.\,\,The similar attempt was made by Kakutani and
Michihiro \cite{Kakutani_1983} (see also a more formal derivation
made later by Parkes \cite{Parkes_1987}).\,\,A general fourth-order
HONLSE for the first-harmonic envelope of surface profile was
derived by Brinch-Nielsen and Jonsson \cite{Brinch-Nielsen_1986} in
coupling with the integral equation for the wave-induced mean flow.
Gramstad and Trulsen \cite{Gramstad_Trulsen_JFM_2011,Gramstad_2014}
derived a fourth-order HONLSE in terms of canonical variables that
preserves the Hamiltonian structure of the surface wave problem.

Sedletsky \cite{SedletskyUJP2003,SedletskyJETP2003} used the
multiple-scale technique to derive a single fourth-order HONLSE for
the first-harmonic envelope of surface profile by introducing an
additional power expansion of the induced mean flow.\,\,This
equation is the direct counterpart of Dysthe's equation written in
terms of the first-harmonic envelope of surface profile
\cite{Trulsen_Dysthe_2000} but for the case of finite depth.
Slunyaev \cite{Slunyaev_2005} confirmed the results obtained in
\cite{SedletskyJETP2003} and extended them to the fifth order.
Grimshaw and Annenkov \cite{Grimshaw_2011} considered a HONLSE for
water wave packets over variable depth.

The deep-water HONLSE in the form of Dysthe's equation was
extensively used in numerical simulations of wave evolution
\cite{Lo_Mei_1985,Akylas_1989,Ablowitz_2000,Ablowitz_2001,Clamond_2006,Slunyaev_2009,Fedele_Dutykh_2011,FD_2012,Slunyaev_PRE_2013}.
However, no such modeling has been performed in the case of finite
depth because of the complexity of equations as compared to the
deep-water limit.\,\,The equation derived in
\cite{SedletskyUJP2003,SedletskyJETP2003} can be used as a good
starting point for the simulations of wave envelope evolution on
finite depth.\,\,The aim of this paper is (i) to rewrite this
equation in dimensionless form suitable for numerical integration
and (ii) to observe the evolution of NLSE solitons taken as initial
waveforms in the case when the HONLSE terms are taken into
consideration for several values of intermediate depth.

This paper is organized as follows.\,\,In Section~\ref{sec:problem},
we write down the fully nonlinear equations of hydrodynamics used as
the starting point in this study.\,\,In Section~\ref{sec:wtrain}, we
formulate the constraints at which the fully nonlinear equations can
be reduced to HONLSE.  Then we briefly outline the multiple-scale
technique used to derive this equation, which is presented in
Section~\ref{sec:mult}.\,\,Next, we introduce dimensionless
coordinate, time, and amplitude to go over to the dimensionless
HONLSE.\,\,As a result, only one dimensionless parameter $kh$
appears in the equation.\,\,The final step is to pass to the
reference frame moving with the group speed of the carrier
wave.\,\,In Section~\ref{sec:num}, we present the results of
numerical simulations and compare the NLSE and HONLSE solutions.
Conclusions are made in Section~\ref{sec:concl}.


\section{Problem Formulation}\label{sec:problem}

\begin{figure}
  \vskip1mm
  \includegraphics[width=\column]{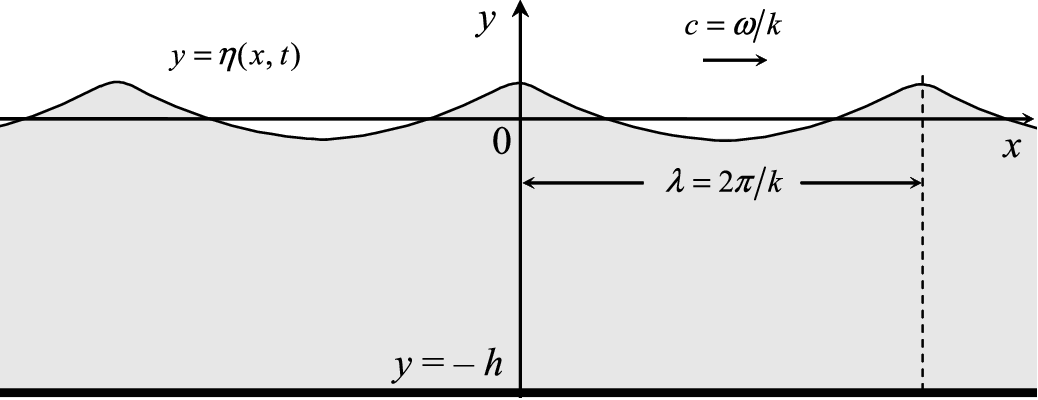}
  \vskip-3mm\caption{Sketch of the physical domain
   occupied by an ideal incompressible fluid of finite depth}
  \label{fig:sketch}
\end{figure}

We consider the dynamics of potential two-di\-men\-sio\-nal waves on
the surface of irrotational, inviscid, and incompressible fluid
under the influence of gravity.\,\,Waves are assumed to propagate
along the horizontal $x$-axis, and the direction of the vertical
$y$-axis is selected opposite to the gravity force.\,\,The fluid is
assumed to be bounded by a solid flat bed $y=-h$ at the bottom and a
free surface $y = \eta(x,\,t)$ at the top
(Fig.~\ref{fig:sketch}).\,\,The atmospheric pressure is assumed to
be constant on the free surface.\,\,Then the evolution of waves and
associated fluid flows is governed by the following system of
equations \cite{Stoker_1992,UJP2013}:
\begin{equation}
\label{eq:Lapl}
 \Phi_{xx}+\Phi_{yy}=0,~~ -\infty<x<\infty,
~~ -h<y<\eta(x,\,t);
\end{equation}\vspace*{-7mm}
\begin{equation}
\label{eq:Dyn}
 \Phi_{t}+\frac{1}{2}\bigl(\Phi_{x}^2+\Phi_y^2\bigr)+g\eta=0,~~~ y=\eta(x,\,t);
\end{equation}\vspace*{-7mm}
 \begin{equation}
\label{eq:Kin}
 \eta_{t}-\Phi_y+\eta_{x}\Phi_{x}=0,~~~ y=\eta(x,\,t);
\end{equation}
\begin{equation}
\label{eq:Bottom}
 \Phi_y=0,~~~  y=-h;
\end{equation}
where $\Phi(x,\,y,\,t)$ is the velocity potential (the velocity is
equal to $\nabla\Phi$), $g$ is the acceleration due to gravity, $t$
is time.\,\,Here, (\ref{eq:Lapl}) is the Laplace equation in the
fluid domain, (\ref{eq:Dyn}) is the dynamical boundary condition
(the so-called Bernoulli or Cauchy--Lagrange integral),
(\ref{eq:Kin}) and (\ref{eq:Bottom}) are the kinematic boundary
conditions (no fluid crosses the free surface and the bottom), the
indices $x$, $y$, and $t$ designate the partial derivatives over the
corresponding variables.\,\,The position of the zero level $y=0$ is
selected such that the Bernoulli constant (the right-hand side of
Eq.~(\ref{eq:Dyn})) is equal to zero.

Consider a modulated wavetrain with carrier frequency $\omega$ and
wavenumber $k$.\,\,In this case, a solution to
Eqs.\,\,(\ref{eq:Lapl})--(\ref{eq:Bottom}) can be looked for in the
form of Fourier series with variable coefficients:
\begin{equation}
\begin{array}{l}
\displaystyle
\!\left(\!\!\begin{array}{c}\Phi(x,\,y,\,t)\\\eta(x,\,t)\end{array}\!\!\right)=
\sum_{n=-\infty}^{\infty}\left(\!\!\begin{array}{c}\Phi_n(x,\,y,\,t)\\\eta_n(x,\,t)\end{array}\!\!\right)
\E^{\I n(kx-\omega t)},\\[5mm]
\displaystyle \eta_{-n}\equiv \eta_n^*,\;\;\Phi_{-n}\equiv
\Phi_n^*,\label{eq:Fourier}
\end{array}\!\!\!\!\!\!\!\!\!\!\!\!\!\!\!\!\!\!\!\!
\end{equation}
where $^*$ stands for complex conjugate (here, we assume the carrier
wave to be symmetric), the functions $\Phi(x,\,y,\,t)$ and
$\eta(x,\,t)$ are assumed to be real by definition.
Substituting \refe{eq:Fourier} in \refe{eq:Lapl}--\refe{eq:Bottom}
and equating the coefficients at the like powers of $\exp(\I
(kx-\omega t))$, one can obtain a system of nonlinear partial
differential equations for the functions $\Phi_n(x,\,y,\,t)$ and
$\eta_n(x,\,t)$.\,\,Linearization of these equations at $n=1$ gives
the dispersion relation for gravity waves:\vspace*{-1mm}
\begin{equation}\label{eq:dispersion}
\omega^2 = gk\tanh(kh).
\end{equation}\vspace*{-6mm}


\section{Slowly Modulated Quasi-Harmonic Wavetrains and
Multiple-Scale Expansions}\label{sec:wtrain}

Generally the system of equations for $\Phi_n(x,\,y,\,t)$ and
$\eta_n(x,\,t)$ is by no means more simple than original
equations.\,\,It can be simplified when solutions are looked for in
a class of functions with narrow spectrum, $|\Delta k| \ll k$.\,\,In
this case, the problem has a formal small parameter $\mu \sim|\Delta
k|\bigl/k$ (quasi-monochromaticity condition), with
$\Phi_n(x,\,y,\,t)$ and $\eta_n(x,\,t)$ being slow functions of $x$
and $t$.\,\,Accordingly, the wave motion can be classified into slow
one and fast one by introducing different time scales and different
spatial scales:
\begin{equation}\label{eq:multiple_scales}
T_n \equiv \mu^n t,\quad X_n \equiv \mu^n x.
\end{equation}
The derivatives with respect to time and coordinate are expanded
into the following series:
\begin{equation}\label{eq:der_expansion}
\frac{\partial}{\partial t} = \sum_{n=0}^{\infty}\mu^n
\frac{\partial}{\partial T_n},\qquad \frac{\partial}{\partial x} =
\sum_{n=0}^{\infty}\mu^n \frac{\partial}{\partial X_n},
\end{equation}
the times $T_n$ and coordinates $X_n$ being assumed to be
independent variables.

When there are no resonances between higher harmonics, the
amplitudes of Fourier coefficients decrease with increasing number
(quasi-harmonicity condition):
\begin{equation}\label{eq:harmonicity}
\eta_n\sim\varepsilon^n A,\quad n\geqslant1,\quad
\eta_0\sim\varepsilon^2 A,\quad\varepsilon < 1,
\end{equation}
where
\begin{equation}\label{eq:A_definition}
\eta_1\equiv\frac{1}{2}\varepsilon A(x,\,t).
\end{equation}
The parameter $\varepsilon$ can be regarded as a formal small
parameter related to the smallness of wave amplitude as compared to
the carrier wavelength $\lambda\equiv\frac{2\pi}{k}$.\,\,In this
case, the unknown functions $\Phi_n(x,\,y,\,t)$ and $\eta_n(x,\,t)$
can be expanded into power series in the formal parameter
$\varepsilon$:
\begin{equation}\label{eq:epsilon_expansion}
\left(\!\!\!\begin{array}{c}\Phi_n(x,\,y,\,t)\\\eta_n(x,\,t)\end{array}\!\!\!\right)=
\sum_{m=1}^{\infty}\varepsilon^m\left(\!\!\begin{array}{c}\Phi_n^{(m)}
(x,\,y,\,t)\\\eta_n^{(m)}(x,\,t)\end{array}\!\!\!\right)\!\!.
\end{equation}
Multiple-scale expansions \refe{eq:der_expansion} and
\refe{eq:epsilon_expansion} allow the functions $\Phi_n(x,\,y,\,t)$
and $\eta_n(x,\,t)$ to be expressed in terms of the first harmonic
envelope $A(x,\,t)$, as described in detail in
\cite{SedletskyUJP2003}.\,\,Note that in the procedure described in
\cite{SedletskyUJP2003} it is essential to set
$\varepsilon\equiv\mu$.

In practice, the quasi-harmonicity condition can be written as
\begin{equation}\label{eq:eta1_smallness}
|k\eta_1|\ll 1,
\end{equation}
and the condition of slow modulation
(quasi-mo\-no\-chro\-ma\-ti\-city) can be formalized as
\begin{equation}\label{eq:Ax_smallness}
\left|\frac{A_x}{kA}\right|\ll 1,
\end{equation}
which follows from differentiating the function $A(x,\,t)
\exp\bigl(\I (kx-\omega t)\bigr)$ with respect to $x$.\,\,With these
conditions satisfied, the original system of equations
(\ref{eq:Lapl})--(\ref{eq:Bottom}) can be reduced to one evolution
equation for the first harmonic envelope $A(x,\,t)$ with the use of
small-amplitude expansions \refe{eq:der_expansion} and
\refe{eq:epsilon_expansion}.


\section{High-Order\\ Nonlinear Schr\"{o}dinger Equation}\label{sec:mult}
\subsection{Equation derived by Sedletsky}

Sedletsky \cite{SedletskyUJP2003,SedletskyJETP2003} used the
above-described multiple-scale procedure to derive the following
HONLSE for the first-harmonic envelope $A(x,\,t)$ (Eq.\,\,(68) in
\cite{SedletskyUJP2003}):
\begin{gather}
\I\left(\!\frac{\partial A}{\partial(\varepsilon
t)}+V_{g}\frac{\partial A}
{\partial(\varepsilon x)}\!\right)+ \nonumber\\
+\,\varepsilon \left(\!\frac{1}{2}\omega'' \frac{\partial^2
A}{\partial(\varepsilon x)^2}
+ \omega k^{2}q_{3}|A|^{2}A\!\right)+ \nonumber\\
+\,\I\varepsilon ^{2}\biggl(\!-\frac{1}{6}\omega'''\frac{\partial^3
A}{\partial(\varepsilon x)^3}
+\omega kQ_{41}|A|^{2}\frac{\partial A}{\partial(\varepsilon x)}\,+\nonumber\\
+\,\omega kQ_{42}A^{2}\frac{\partial A^{*}}{\partial(\varepsilon
x)}\!\biggr) =0\;\;[\text{m}/\text{s}].\label{eq:Sedletsky}
\end{gather}
As compared to the standard NLSE, this equation takes into account
additional nonlinear and dispersive terms of order
$\O(\varepsilon^2)$.\,\,Equation~\refe{eq:Sedletsky} was later
re-derived by Slunyaev \cite{Slunyaev_2005}, who confirmed the
symbolic computations presented in
\cite{SedletskyUJP2003,SedletskyJETP2003} and extended them to the
$\O(\varepsilon^3)$ order.\,\,Here, we restrict our attention to the
original equation \refe{eq:Sedletsky}.\,\,The parameters of this
equation are given by
\begin{subequations}
\begin{equation}
\omega =\left(gk\sigma\right)^{1/2},\quad \sigma \equiv \tanh(kh),
\end{equation}\vspace*{-9mm}
\[
\omega'=\frac{\partial \omega}{\partial k} \equiv V_g =
\frac{\omega}{2k}\left(\!1+\frac{2k h}{\sinh(2kh)}\!\right)
   =
\]\vspace*{-7mm}
\begin{equation}
   =\frac{\omega}{2k}\left(\!1+\frac{1-\sigma^2}{\sigma}kh\!\right)\!\!,
\end{equation}\vspace*{-7mm}
\[
\omega''=\frac{\partial^2 \omega}{\partial k^2} =
\frac{\omega}{4k^2\sigma^2}\Bigl(\!\left(\sigma^2-1\right)\left(3\sigma^2+1\right)k^2
h^2\,-
\]\vspace*{-7mm}
\begin{equation}
 -\,2\sigma\left(\sigma^2-1\right)kh-\sigma^2\!\Bigr)\!,
\end{equation}
\[
\omega'''=\frac{\partial^3 \omega}{\partial k^3} =
-\frac{\omega}{8k^3\sigma^3} \Bigl(\!\left(\sigma^2-1\right)\times
\]\vspace*{-7mm}
\[
\times \left(15\sigma^4\!-2\sigma^2+3\right)k^3
h^3\!-3\sigma\left(\sigma^2\!-1\right)\!\left(3\sigma^2\!+1\right)k^2
h^2\,-
\]\vspace*{-7mm}
\begin{equation}
-\, 3\sigma^2\left(\sigma^2-1\right)kh-3\sigma^3\!\Bigr)\!,
\end{equation}\vspace*{-9mm}
\[
q_3=-\frac{1}{16\sigma^4\nu}\Bigl(\!\left(\sigma^2-1\right)^2
\left(9\sigma^4-10\sigma^2+9\right)k^2 h^2\,+
\]\vspace*{-7mm}
\[
+\, 2\sigma\left(3\sigma^6-23\sigma^4+13\sigma^2-9\right)kh\,-
\]\vspace*{-7mm}
\begin{equation}
  -\,\sigma^2
   \left(7\sigma^4-38\sigma^2-9\right)\!\Bigr)\!,
\end{equation}\vspace*{-9mm}
\[
Q_{41}=\frac{1}{32\sigma^5\nu^2}\Bigl(\!\left(\sigma^2-1\right)^5\times\]\vspace*{-7mm}
\[\times
\left(3\sigma^6-20\sigma^4-21\sigma^2+54\right)k^5 h^5\,-
\]\vspace*{-8mm}
\[
 -\,\sigma\left(\sigma^2\!-1\right)^3\bigl(11\sigma^8\!-99\sigma^6\!-61\sigma^4\!+
7\sigma^2\!+270\bigr)k^4 h^4\,+
\]\vspace*{-8mm}
\[
+\,2\sigma^2\left(\sigma^2-1\right)\bigl(7\sigma^{10}-58\sigma^8+38\sigma^6+
52\sigma^4\,-
\]\vspace*{-8mm}
\[
-\,181\sigma^2+270\!\bigr) k^3
h^3-2\sigma^3\bigl(3\sigma^{10}+18\sigma^8-146\sigma^6\,-
\]\vspace*{-9mm}
\[
-\,172\sigma^4+183\sigma^2-270\bigr)k^2
h^2-\sigma^4\bigl(\sigma^8-109\sigma^6+517\sigma^4\,+
\]\vspace*{-9mm}
\begin{equation}
+\,217\sigma^2 + 270\bigr)kh
    +\sigma^5\left(\sigma^6-40\sigma^4+193\sigma^2+54\right)\!\Bigr)+\Delta,
\end{equation}\vspace*{-12mm}
\[
Q_{42}=\frac{1}{32\sigma^5\nu^2}\Bigl(\!-\left(\sigma^2-1\right)^5\times\]\vspace*{-7mm}
\[\times\left(3\sigma^6+7
\sigma^4-11\sigma^2+9\right)k^5 h^5\,+
\]\vspace*{-9mm}
\[
+\,\sigma\left(\sigma^2\!-1\right)^3\left(11\sigma^8\!-48\sigma^6\!+66\sigma^4\!+8\sigma^2\!+27\right)
 k^4 h^4\,-
\]\vspace*{-9mm}
\[-\,2\sigma^2\left(\sigma^2-1\right)\bigl(7\sigma^{10}-79\sigma^8+282
\sigma^6\,-
\]\vspace*{-9mm}
\[
-\, 154\sigma^4-\sigma^2+9\bigr)k^3 h^3+2\sigma^3
\bigl(3\sigma^{10}-63\sigma^8+314\sigma^6\,-
\]\vspace*{-9mm}
\[
-\,218\sigma^4+19\sigma^2+9 \bigr)k^2
h^2+\sigma^4\bigl(\sigma^8+20\sigma^6-158\sigma^4\,-
\]\vspace*{-8mm}
\begin{equation}
-\,28\sigma^2-27\bigr) kh
-\sigma^5\left(\sigma^6-7\sigma^4+7\sigma^2-9\right)\!\Bigr)-\Delta,
\end{equation}\vspace*{-7mm}
\begin{equation}
\nu =\left(\sigma^2-1\right)^2 k^2 h^2 -
2\sigma\left(\sigma^2+1\right)kh + \sigma^2.
\end{equation}
\end{subequations}
The quantity $V_g$ is the wave group speed.\,\,The parameter
$\Delta$ is the correction introduced by
Slunyaev~\cite{Slunyaev_2005} to the coefficients derived in
\cite{SedletskyUJP2003,SedletskyJETP2003}.\,\,This correction is
negligible at $kh\gtrsim 1$ (see Appendix~\ref{app:slu}), and we
ignore it by keeping $\Delta = 0$.

The free-surface displacement is expressed in terms of $A$ as
\[
\eta = \varepsilon^2 \eta_0 + \varepsilon \Re\bigl( A \E^{\I
(kx-\omega t)}\bigr)\,+
\]\vspace*{-7mm}
\begin{equation}\label{Eq:eta}
+\,\varepsilon^2\,2\Re\bigl( \eta_2\, \E^{2\I(kx-\omega t)}\bigr) +
\O(\varepsilon^3),
\end{equation}
where $\Re\{\cdot\}$ stands for the real part of a complex-valued
function.\,\,Here, $\eta_{0}$ and $\eta_{2}$ are defined
as\pagebreak[0]
\begin{subequations}
\begin{equation}
\eta_0 = \frac{\sigma+2\left(1-\sigma^2\right)kh}{\nu}\,k|A|^2,
\end{equation}\vspace*{-7mm}
\begin{equation}
\eta_2 = \frac{3-\sigma^2}{8\sigma^3}\,kA^2.
\end{equation}
\end{subequations}
The corresponding velocity potential is written as
\[
\Phi = \varepsilon \Phi_0 + \varepsilon\,2\Re\bigl(\Phi_1\, \E^{\I
(kx-\omega t)}\bigr)\,+
\]\vspace*{-7mm}
\begin{equation}\label{Eq:Phi}
+\,\varepsilon^2\,2\Re\bigl(\Phi_2\, \E^{2\I(kx-\omega t)}\bigr) +
\O(\varepsilon^3),
\end{equation}
where\vspace*{-3mm}
\begin{subequations}
\[
\Phi_1 = \frac{\omega}{2k\sigma}\Biggl(\!\biggl(\!\frac{\partial
A}{\partial x} \biggl(h\sigma+\frac{V_g}{\omega}\biggr)-\I
A\!\biggr)\frac{\cosh\bigl(k(y+h)\bigr)}{\cosh(kh)}\,-
\]\vspace*{-5mm}
\begin{equation}
-\,(y+h)\frac{\partial A}{\partial
x}\,\frac{\sinh\bigl(k(y+h)\bigr)}{\cosh(kh)}\!\Biggr)\!,
\end{equation}\vspace*{-5mm}
 \begin{equation}
\Phi_2 =
3\I\omega\,\frac{(\sigma^4-1)}{16\sigma^4}\,\frac{\cosh\bigl(2k(y+h)\bigr)}{\cosh(2kh)}\,A^2.
\end{equation}
\end{subequations}
The term $\Phi_0$ describes the wave-induced mean flow and is
expressed implicitly in terms of its derivatives
\begin{subequations}
\begin{equation}
\frac{\partial \Phi_0}{\partial x} = \varepsilon\frac{\omega k
\gamma_1}{2\sigma\nu}|A|^2+
 \I\varepsilon\frac{\omega\gamma_2}{8\sigma^2\nu^2}\left(\!A\frac{\partial A^*}
 {\partial x}-A^*\frac{\partial A}{\partial x}\!\right)\!\!,
\end{equation}\vspace*{-7mm}
\begin{equation}
\frac{\partial \Phi_0}{\partial t} = -V_g \frac{\partial
\Phi_0}{\partial x},
\end{equation}
\end{subequations}
where
\begin{subequations}
\begin{equation}
\gamma_1 =\bigl(\sigma^2-1\bigr)^2 kh -
\sigma\bigl(\sigma^2-5\bigr),
\end{equation}\vspace*{-6mm}
\[
\gamma_2 =\bigl(\sigma^2-1\bigr)^5 k^4h^4 +
4\sigma\bigl(\sigma^2-1\bigr)^2 \bigl(13\sigma^2+3\bigr)k^3h^3\,-
\]\vspace*{-6mm}
\[
-\,2\sigma^2\bigl(\sigma^2-1\bigr)\bigl(3\sigma^4+32\sigma^2-3\bigr)k^2h^2\,+
\]\vspace*{-6mm}
\begin{equation}
+\,4\sigma^3\bigl(2\sigma^4-\sigma^2-5\bigr)kh -
3\sigma^4\bigl(\sigma^2-5\bigr).
\end{equation}
\end{subequations}
Functions \refe{Eq:eta} and \refe{Eq:Phi} define an approximate
solution to the original system of equations
(\ref{eq:Lapl})--(\ref{eq:Bottom}) in terms of the first-harmonic
envelope $A$, which is found from Eq.~\refe{eq:Sedletsky}.

\subsection{Dimensionless form}
\begin{figure}[b!]
 \vskip-3mm
 \includegraphics[width=\column]{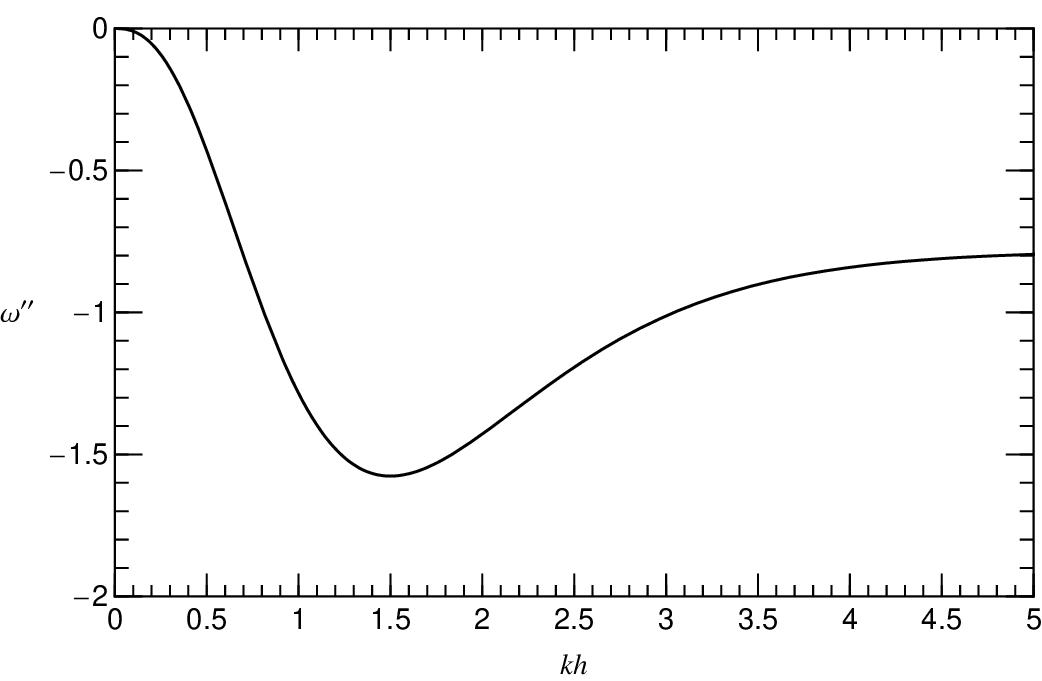}
 \vskip-3mm\caption{\label{fig:w2p} $\omega''$ in $\mathrm{m}^2/\mathrm{s}$
  as a function of $h$ at $k=1$ and
$g=9.8\,\mathrm{m}/\mathrm{s}^2$}\vspace*{1.5mm}
\end{figure}
\begin{figure*}[t]
 \vskip1mm
\includegraphics[width=11.5cm]{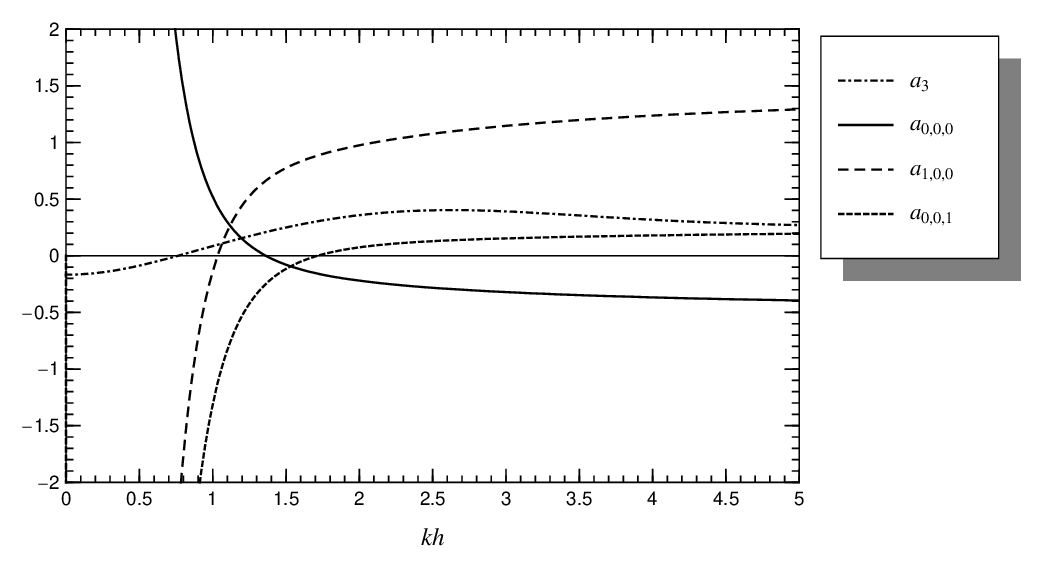}
 \vskip-2mm\parbox{12cm}{\caption{\label{fig:an} Normalized coefficients of HONLSE as
functions of $kh$\hspace*{3.7cm}}}\vskip2mm
\end{figure*}

Introduce the following dimensionless time, coordinate, and amplitude:
\begin{equation}
\tau = \beta t,\quad \chi = kx,\quad u = \alpha^{-1} \varepsilon A,
\end{equation}
$\alpha$ and $\beta$ being the parameters to be determined.\,\,The
relationship between the old and new derivatives is
\begin{equation*}
\frac{\partial}{\partial x} = k\frac{\partial}{\partial \chi},\quad
\frac{\partial}{\partial t} = \beta\frac{\partial}{\partial \tau}.
\end{equation*}
Then Eq. (\ref{eq:Sedletsky}) is transformed to
\[
\I\alpha\left(\beta u_\tau + k V_g u_\chi\right) +\alpha
\left(\!\frac{1}{2}\omega'' k^2 u_{\chi\chi}+\omega k^{2}q_{3}
|\alpha|^2 |u|^{2}u\!\right)+
\]\vspace*{-5mm}
\[
+\,\I\alpha\Bigl(-\frac{1}{6}\omega''' k^3 u_{\chi\chi\chi}+\omega
k^2 Q_{41}|\alpha|^2 u_{\chi}|u|^{2}\,+
\]\vspace*{-5mm}
\[
+\,\omega k^2 Q_{42}|\alpha|^2 u^{2}u^{*}_{\chi}\Bigr)
=0\;\;[\text{m}/\text{s}].
\]
Here, the indices $\chi$ and $\tau$ designate the partial
derivatives with respect to the corresponding
va\-riab\-les.\,\,Ta\-king into account that $\omega''<0$ at all
$h>0$ (Fig.\,\,\ref{fig:w2p}), divide this equation by $\omega'' k^2
\alpha$ so that
\[
\I\left(\!\frac{\beta}{\omega'' k^2} u_\tau + \frac{V_g}{\omega'' k}
u_\chi\!\right)
 +\left(\!\frac{1}{2} u_{\chi\chi}+\frac{\omega}{\omega''}|\alpha|^2 q_{3}
 |u|^{2}u\!\right)+
\]\vspace*{-6mm}
\[
+\,\I\Bigl(\!-\frac{1}{6}\frac{\omega''' k}{\omega''}
u_{\chi\chi\chi}+\frac{\omega}{\omega''}|\alpha|^2 Q_{41}
u_{\chi}|u|^{2}\,+\]\vspace*{-6mm}
\[ +\,\frac{\omega}{\omega''}|\alpha|^2 Q_{42}
u^{2}u^{*}_{\chi}\!\Bigr) =0
\]
and select the values of $\alpha$ and $\beta$ as
\begin{equation}
|\alpha|^2 = -\frac{\omega''}{\omega}>0,\quad \beta = -\omega''k^2>0.
\end{equation}
Thus, Eq.\,\,(\ref{eq:Sedletsky}) takes the dimensionless form
\[
\I \left(\!u_\tau - \frac{V_g}{\omega'' k} u_\chi\!\right)
-\frac{1}{2}u_{\chi\chi}+q_3|u|^2 u\,+
\]\vspace*{-6mm}
\[
 +\,
\I\left(\!\frac{1}{6}\frac{\omega'''k}{\omega''}u_{\chi\chi\chi}+
Q_{41}u_{\chi}|u|^{2}+Q_{42}u^{2}u^{*}_{\chi}\!\right)=0
\]
or, equivalently,
\begin{subequations}\label{Eq68_u}
\[
\I \bigl(u_\tau + a_1 u_\chi\bigr) - a_2
u_{\chi\chi}+a_{0,\,0,\,0}|u|^2 u+ \I\Bigl(-a_3
u_{\chi\chi\chi}\,+\]\vspace*{-7mm}
\begin{equation}
+\, a_{1,\,0,\,0}u_{\chi}|u|^{2} +
a_{0,\,0,\,1}u^{2}u^{*}_{\chi}\Bigr)=0,
\end{equation}
which finally yields
\[
u_\tau = - a_1 u_\chi - \I a_2 u_{\chi\chi}+\I a_{0,\,0,\,0}|u|^2
u\,+
\]\vspace*{-7mm}
\begin{equation}
+\, \Bigl(a_3 u_{\chi\chi\chi} - a_{1,\,0,\,0}u_{\chi}|u|^{2} -
a_{0,\,0,\,1}u^{2}u^{*}_{\chi}\Bigr)\!,
\end{equation}
\end{subequations}
where we used the unified notation introduced by Lukomsky and
Gandzha \cite{ujp09en}. Here, the coefficients
\begin{equation}
\begin{array}{l}
\displaystyle a_1 = -\frac{V_g}{\omega'' k} =
-\frac{2}{\upsilon}\Bigl(\!\sigma^2+\sigma\left(1-\sigma^2\right)kh\!\Bigr)>0,\\[3mm]
\displaystyle a_2 = \frac{1}{2},\\[3mm]
\displaystyle a_3 \equiv -\frac{1}{6}\frac{\omega'''k}{\omega''}=\\[3mm]
\displaystyle  =
\frac{1}{12\sigma\upsilon}\Bigl(\!\left(\sigma^2-1\right)
\left(15\sigma^4-2\sigma^2+3\right)k^3 h^3\,-\\[3mm]
 \displaystyle -\,3\sigma\left(\sigma^2-1\right)\left(3\sigma^2+1\right)k^2 h^2\,-\\[1mm]
\displaystyle -\,3\sigma^2\left(\sigma^2-1\right)kh-3\sigma^3\!\Bigr)\!,\\[1mm]
\displaystyle a_{0,\,0,\,0} \equiv q_3,\quad a_{1,\,0,\,0} \equiv
Q_{41},\quad a_{0,\,0,\,1}
 \equiv Q_{42}, \\[1mm]
\displaystyle \upsilon =
\left(\sigma^2-1\right)\left(3\sigma^2+1\right)k^2
h^2-2\sigma\left(\sigma^2-1\right)kh-\sigma^2
\end{array}\!\!\!\!\!\!\!\!\!\!\!\!\!\!\!\!\!\!\!\!\!\!\!\!
\end{equation}
are all real and depend on one dimensionless parameter
$kh$.\,\,Their behavior as functions of $kh$ is shown in
Fig.\,\,\ref{fig:an}.\,\,It can be seen that
Eq.\,\,(\ref{eq:Sedletsky}) is valid at \mbox{$kh \gtrsim 1$,} where
the coefficients $a_{0,\,0,\,0}$, $a_{1,\,0,\,0}$, and
$a_{0,\,0,\,1}$ do not diverge.\,\,At smaller depths, the
Kor\-te\-weg--de Vries equation and its generalizations
\cite{Johnson_2002,Kharif_Enpeli_2003} should be used.\,\,On the
other hand, at large $kh$, the infinite-depth limit (Dysthe's
equation) should be used.\,\,Indeed, the following asymptotics are
easily obtained at $kh\rightarrow\infty$:
\begin{equation}
a_3 = \frac{1}{4},\; a_{0,\,0,\,0} = -\frac{1}{2},\; a_{1,\,0,\,0} =
\frac{3}{2},\; a_{0,\,0,\,1} = \frac{1}{4}.\!\!\!\!
\end{equation}
They coincide with the corresponding coefficients of Dysthe's
equation \cite{Trulsen_Dysthe_2000}, except for the term including
the wave-induced mean flow, which cannot be exp\-li\-cit\-ly
reconstructed from Eq.\,\,\refe{Eq68_u} because of the additional
power expansion of the wave-induced mean flow made to derive
Eq.\,\,\refe{eq:Sedletsky}.\,\,However, this term can be
reconstructed from the equations gene\-ra\-ting
Eq.\,\,\refe{eq:Sedletsky}, at the stage when the wave-induced mean
flow has not been excluded from the equation for $A$ yet
\cite{SedletskyUJP2003}.\,\,Taking into account these constraints,
we will restrict our attention to the following range of
intermediate depths:\vspace*{-2mm}
\begin{equation}
1 < kh < 5.
\end{equation}

\subsection{Moving reference frame}

Equation~\refe{Eq68_u} can be rewritten in the form without the
$u_\chi$ term.\,\,To this end, let us proceed to the reference frame
moving with speed $a_1$ (dimensionless group speed):
\begin{equation}
\xi = \chi-a_1 \tau,\quad T = \tau.
\end{equation}
The relationship between the derivatives in new and old variables is
given by the formulas
\[
\frac{\partial}{\partial \chi} = \frac{\partial \xi}{\partial
\chi}\frac{\partial}{\partial \xi} +
  \frac{\partial T}{\partial \chi}\frac{\partial}{\partial T} = \frac{\partial}{\partial \xi},
  \]\vspace*{-5mm}
  \[
\frac{\partial}{\partial \tau}= \frac{\partial \xi}{\partial
\tau}\frac{\partial}{\partial \xi} +
  \frac{\partial T}{\partial \tau}\frac{\partial}{\partial T}
  = -a_1\frac{\partial}{\partial \xi} + \frac{\partial}{\partial T},
\]
so that
\[
u_\tau = - \I a_2 u_{\xi\xi}+\I a_{0,\,0,\,0}|u|^2 u\,+
\]
\begin{equation}\label{Eq_u}
 +\, \Bigl(\!a_3 u_{\xi\xi\xi} - a_{1,\,0,\,0}u_{\xi}|u|^{2} - a_{0,\,0,\,1}u^{2}u^{*}_{\xi}\!\Bigr)\!.
\end{equation}
This is our target equation for numerical simulations.\,\,It
possesses the integral of motion
\begin{equation}\label{Eq:I0}
I_0(\tau) = \int\limits_{-\infty}^{\infty}|u(\xi,\,\tau)|^{2}\D\xi=\mathrm{const},
\end{equation}
which expresses the conservation of wave action.\,\,The derivation
of this conservation law is given in Appendix~\ref{app:cl}.\,\,It
allows one to trace the relative numerical error of simulations:
\begin{equation}\label{Eq:ErI0}
\mathrm{Er}(I_0) = \frac{|I_0(\tau)-I_0(0)|}{I_0(0)}.
\end{equation}

\begin{figure}[t]
 \vskip1mm
 \includegraphics[width=\column]{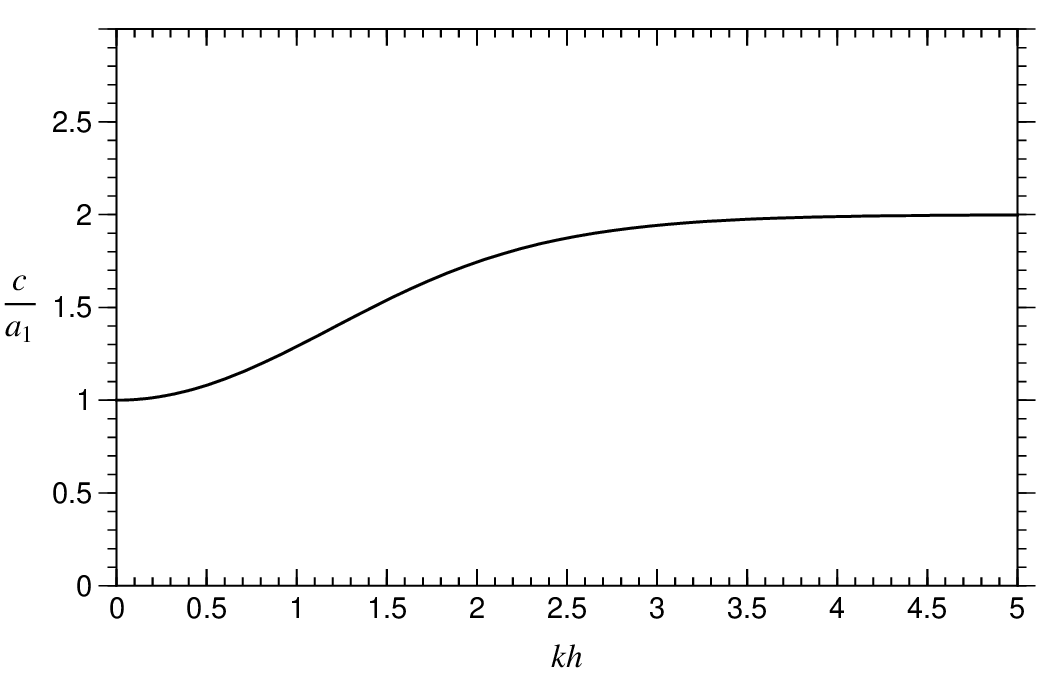}
 \vskip-3mm\caption{\label{fig:ca1}Ratio between the
  dimensionless phase and group speeds as a function of $kh$}
\end{figure}

Of particular interest is to reveal any relationship of
Eq.\,\,\refe{Eq_u} to other HONLSEs derived in different
contexts.\,\,In Appendix~\ref{app:ss}, we consider one such equation
(the Sasa--Satsuma equation) and prove that Eq.\,\,\refe{Eq_u}
cannot be reduced to it at any $kh$.

\subsection{Dimensionless free surface\\ displacement and velocity potential}

The dimensionless free surface displacement is expressed in terms of
$u$ as follows:
\begin{equation}\label{Eq:zeta}
\zeta\equiv k\eta =\alpha_0|u|^2 + \alpha_1\Re\bigl(
u\,\E^{\I\theta}\bigr)
  + 2\alpha_2\Re\bigl( u^2\, \E^{2\I\theta}\bigr),
\end{equation}\vspace*{-7mm}
\begin{equation*}
\alpha_0 =\frac{\sigma+2\left(1-\sigma^2\right)kh}{c\nu},
\quad\alpha_1 = \frac{1}{\sqrt{c}},\quad
\alpha_2=\frac{3-\sigma^2}{8c\sigma^3},
\end{equation*}
where
\begin{equation}
\theta = kx - \omega t = \chi - c\tau = \xi + \left(a_1 -c\right)\tau
\end{equation}
is the wave phase and
\begin{equation}
c = \frac{1}{|\alpha|^2k^2} = -\frac{4\sigma^2}{\upsilon}
\end{equation}
is the dimensionless phase speed.\,\,Figure~\ref{fig:ca1} shows the
ratio of the phase speed $c$ to the group speed $a_1$ as a function
of $kh$.\,\,This ratio is equal to unity at $kh \rightarrow 0$, and
it is twice as large at $kh \rightarrow \infty$, in full conformity
with the classical water wave theory \cite{UJP2013}.\,\,The wave
envelope is written as
\begin{equation}
\left[\zeta\right]_{\mathrm{envelope}} = \alpha_1|u| + \left(\alpha_0 + 2\alpha_2\right)|u|^2.
\end{equation}

The corresponding dimensionless velocity potential is expressed as
\[
\varphi\equiv -\frac{1}{\omega''}\Phi =\varphi_0 +
2\Re\bigl(\varphi_1\,\E^{\I\theta}\bigr) \,+
\]\vspace*{-7mm}
\begin{equation}\label{Eq:varphi}
+\,2\Re\bigl(\varphi_2\,\E^{2\I\theta}\bigr),
\end{equation}\vspace*{-7mm}
\[
\left(\varphi_0\right)_{\xi} =\frac{\gamma_1}{2\sigma\nu}|u|^2+
 \frac{\I\gamma_2}{8\sigma^2\nu^2}\left(uu^*_{\xi}-u^*u_{\xi}\right)\!,
 \]\vspace*{-7mm}
 \[
\left(\varphi_0\right)_{\tau} =-a_1 \left(\varphi_0\right)_{\xi},
\]\vspace*{-7mm}
\[
\varphi_1 = \frac{\sqrt{c}}{2\sigma}\Biggl(\!\!\biggl(\!-\I u +
\frac{\bigl(\sigma^2+1\bigr)kh+\sigma}{2\sigma}\,u_{\xi}\!\biggr)\times
\]\vspace*{-7mm}
\[
 \times\,\frac{\cosh(z+kh)}{\cosh(kh)}-(z+kh)\frac{\sinh(z+kh)}{\cosh(kh)}\,u_{\xi}\!\Biggr)\!,
 \]
 \[
\varphi_2
=\frac{3\I\bigl(\sigma^4-1\bigr)}{16\sigma^4}\,\frac{\cosh(2(z+kh)\bigr)}{\cosh(2kh)}\,u^2,
\]
where $z\equiv ky$ is the dimensionless vertical coordinate.\,\,The
quasi-harmonicity condition is written as
\begin{equation}\label{eq:QH}
\frac{|u|}{\sqrt{c}}\ll 1,
\end{equation}
and the quasi-monochromaticity condition is
\begin{equation}\label{eq:QM}
\left|\frac{u_{\xi}}{u}\right|\ll 1.
\end{equation}
Finally, the original equations of hydrodynamics can be written in
the following dimensionless form:
\begin{equation}\label{eq:Lapl_dim}
 \varphi_{\xi\xi}+\varphi_{zz}=0,
 ~-\infty<\xi<\infty,~
\displaystyle-kh<z<\zeta(\xi,\,\tau);
\end{equation}\vspace*{-7mm}
\begin{equation}\label{eq:Dyn_dim}
 \varphi_{\tau}+\frac{1}{2}\bigl(\varphi_{\xi}^2+\varphi_z^2\bigr)
 +\frac{c^2}{\sigma}\,\zeta=0,~~~z=\zeta(\xi,\,\tau);
\end{equation}\vspace*{-7mm}
\begin{equation}
\label{eq:Kin_dim}
 \zeta_{\,\tau}-\varphi_z+\zeta_{\,\xi}\,\varphi_{\xi}=0,~~~ z=\zeta(\xi,\,\tau);
\end{equation}\vspace*{-7mm}
\begin{equation}
\label{eq:Bottom_dim}
 \varphi_z=0,~~~z=-kh.
\end{equation}


\section{Numerical Simulations}\label{sec:num}

\begin{figure}[t]
 \vskip1mm
 \includegraphics[width=\column]{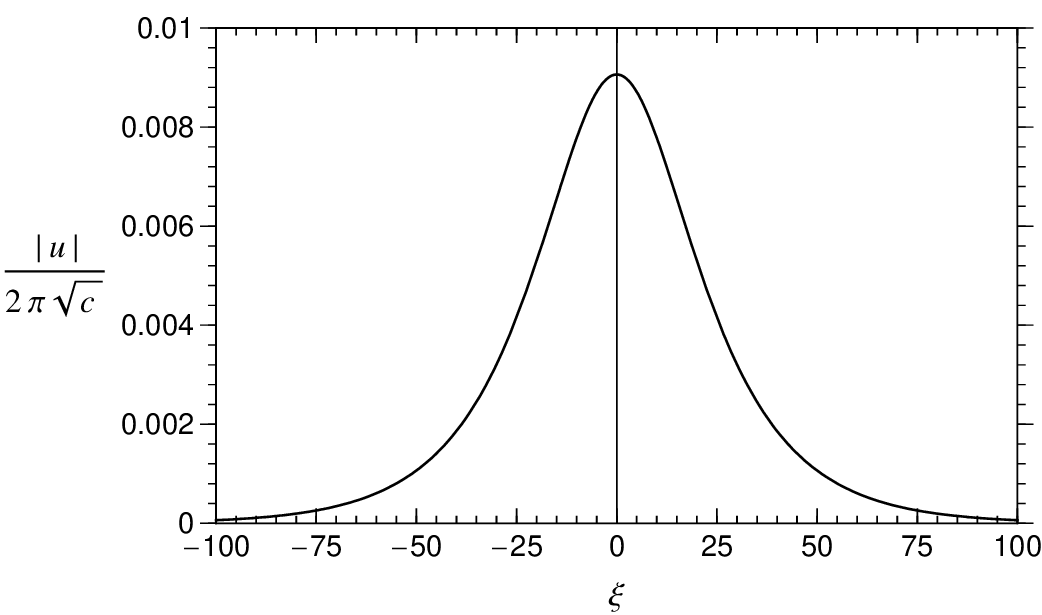}
 \vskip-3mm\caption{\label{fig:QH}Testing the quasi-harmonicity condition \refe{eq:QH} for \mbox{$kh = 3$}}\vspace*{-2mm}
\end{figure}

\begin{figure}[t]
 \vskip3mm
 \includegraphics[width=8cm]{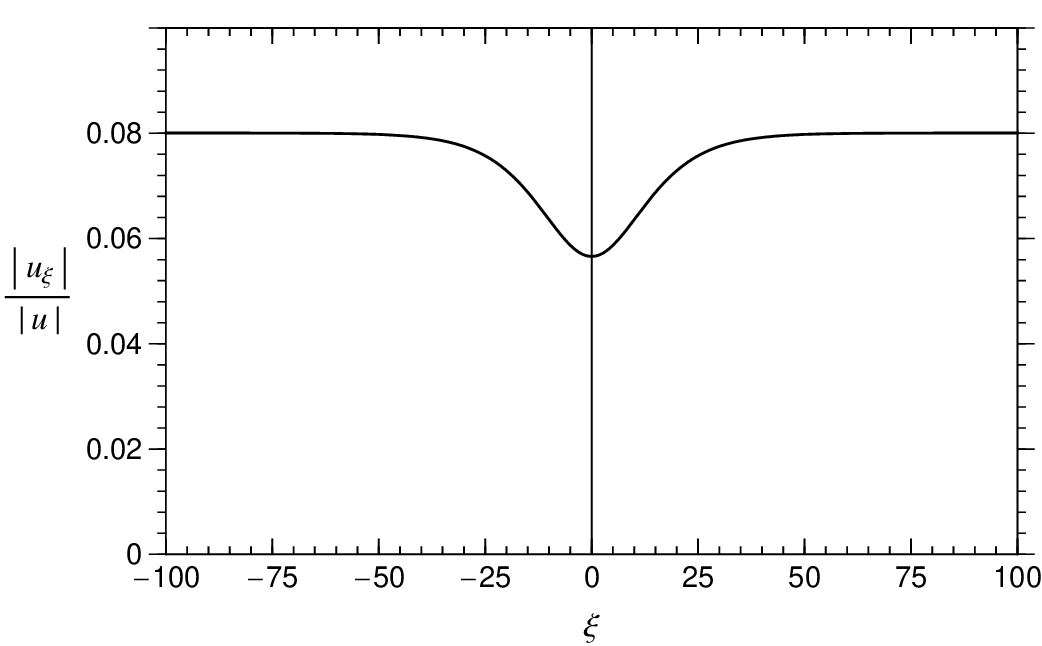}
 \vskip-3mm\caption{\label{fig:QM}Testing the quasi-monochromaticity condition \refe{eq:QM} for $kh =
 3$}\vspace*{-2mm}
\end{figure}

In this section, we adopt the split-step Fourier (SSF) technique
described in Appendix~\ref{app:ssf} to compute solutions to HONLSE
\refe{Eq_u}.\,\,To test the accuracy of our numerical scheme, we
start from classical NLSE \refe{eq:NLSE} written in terms of the
coordinate $\xi$.\,\,At $a_{0,\,0,\,0} < 0$ ($kh\gtrsim1.363$), it
has an exact one-soliton solution~\cite{Zakharov_1971}:
\begin{equation}\label{soliton}
u(\xi,\,\tau)=\frac{u_0\exp\left(\I\kappa\xi-\I\Omega\tau\right)}{\cosh\bigl(K(\xi-\xi_0-V\tau)\bigr)},
\end{equation}\vspace*{-7mm}
\[
\Omega = \left(K^2-\kappa^2\right)a_2,\; V = -2\kappa a_2,\; K =
|u_0|\sqrt{-\frac{a_{0,\,0,\,0}}{2a_2}},
\]\vspace*{-7mm}
\[
u_0\in\mathds{C},\quad \kappa,\;\xi_0\in\mathds{R}.
\]
Here, $V$ is the soliton speed, $u_0$ is the complex amplitude,
$\kappa$ and $\Omega$ are the soliton's wavenumber and frequency,
and $\xi_0$ is the soliton's initial position.\,\,The amplitude
$u_0$ and wavenumber $\kappa$ should be selected such that the
qua\-si-har\-mo\-ni\-ci\-ty and qua\-si-mo\-no\-chro\-ma\-ti\-ci\-ty
conditions \refe{eq:QH}, \refe{eq:QM} hold true.\,\,In practice,
these conditions mean that the soliton amplitude and wavenumber
should be small:
\begin{equation*}
|u_0| \ll 1, \quad \kappa \ll 1.
\end{equation*}
In this study, we restrict our attention by the following choice of
parameters:
\begin{equation}
u_0=0.1,\quad \kappa = -K\;(\Rightarrow\; \Omega=0),\quad \xi_0 = 0.
\end{equation}
Figures \ref{fig:QH} and \ref{fig:QM} demonstrate that constraints
\refe{eq:QH} and \refe{eq:QM} are readily satisfied in this
case.\,\,Note that at $\kappa <0$ we have $V>0$. In this case,
solitons move from left to right with speed exceeding the carrier
group speed.

\begin{figure*}[t]
 \vskip1mm
 \includegraphics[width=11.5cm]{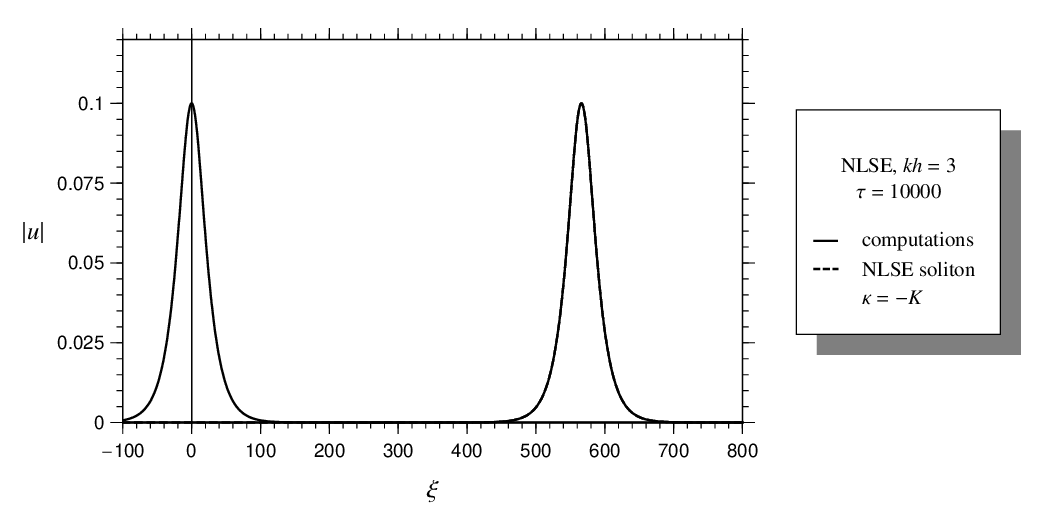}
 \vskip-3mm\parbox{11.5cm}{\caption{\label{fig:NLS} Evolution of one-soliton solution (\ref{soliton}) to NLSE (\ref{eq:NLSE}) at $kh=3$.\,\,SSF parameters:
 $\Delta\tau = 1,\; \Delta\xi = 2,\; \xi\in[-1000,\,1000); \; V \approx
 0.0566$}}
\end{figure*}

Figure \ref{fig:NLS} shows a soliton computed for $kh = 3$ using
analytical formula \refe{soliton} for the initial moment $\tau = 0$
and moment $\tau = 10 000$.\,\,The same soliton was taken as the
initial condition for the simulation with the SSF technique.\,\,The
deviation from the exact solution is seen to be
negligible.\,\,Indeed, the numerical error estimated with formula
\refe{Eq:ErI0} is
\[
S^{(2)}|_{\tau=10000}:~~ \mathrm{Er}(I_0) = 1.3\times10^{-10}\%,
\]\vspace*{-8mm}
\[
 \Delta_{\mathrm{rms}}(u_{\mathrm{exact}},\,u_{\mathrm{comp}})=1.0\times10^{-4}\%,
 \]\vspace*{-8mm}
 \[
S^{(4)}|_{\tau=10000}:~~\mathrm{Er}(I_0) = 2.3\times10^{-10}\%,
\]\vspace*{-8mm}
\[
\Delta_{\mathrm{rms}}(u_{\mathrm{exact}},\,u_{\mathrm{comp}})=3.5\times10^{-9}\%,
\]
where $S^{(2)}$ and $S^{(4)}$ designate the order of the SSF technique
adopted for calculation (see Appendix~\ref{app:ssf}) and
\begin{equation*}
\Delta_{\mathrm{rms}}(u,\,g)(\tau)=\sqrt{\frac{\int_{-\infty}^{\infty}\left(|u(\xi,\,\tau)|-|g(\xi,\,\tau)|\right)^2\D\xi}{I_0}}
\end{equation*}
is the relative r.m.s.~deviation between two functions.\,\,Thus, our
numerical scheme reproduces the exact one-soliton solution to NLSE
with high accuracy.

\begin{figure*}[!]
 \vskip2mm
 \includegraphics[width=11.5cm]{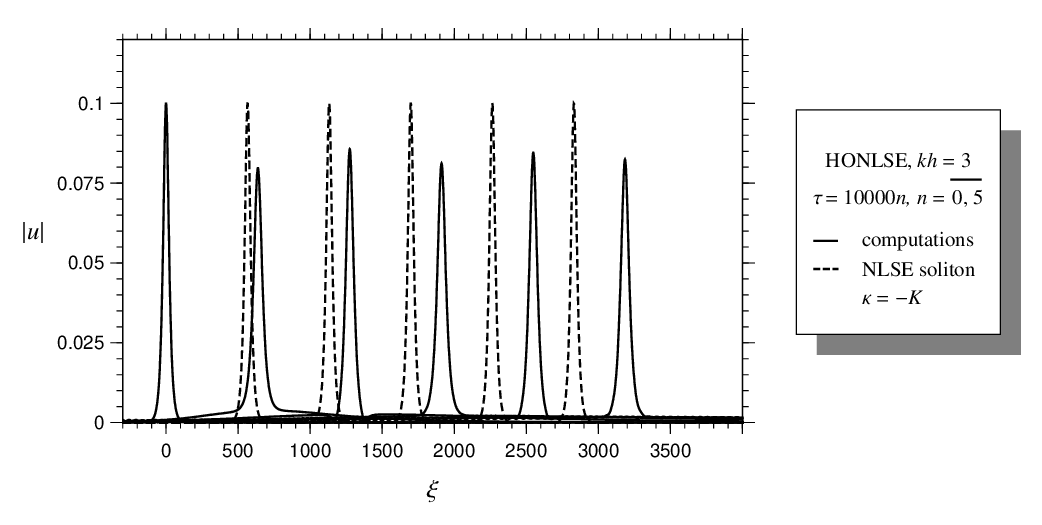}
\vskip-3mm\parbox{11.5cm}{\caption{\label{fig:HONLSE_kh3} Evolution
of one-soliton waveform (\ref{soliton}) taken as the initial
condition in HONLSE \refe{Eq_u} at $kh=3$.\,\,SSF parameters:
$\Delta\tau = 0.5,\; \Delta\xi =
2,\;\xi\in[-4000,\,4000)$.\,\,Accuracy: $S^{(2)}|_{\tau=50000}$:
$\mathrm{Er}(I_0) = 0.050$\%}}\vspace*{-1mm}
\end{figure*}

\begin{figure}[!]
 \vskip2mm
 \includegraphics[width=\column]{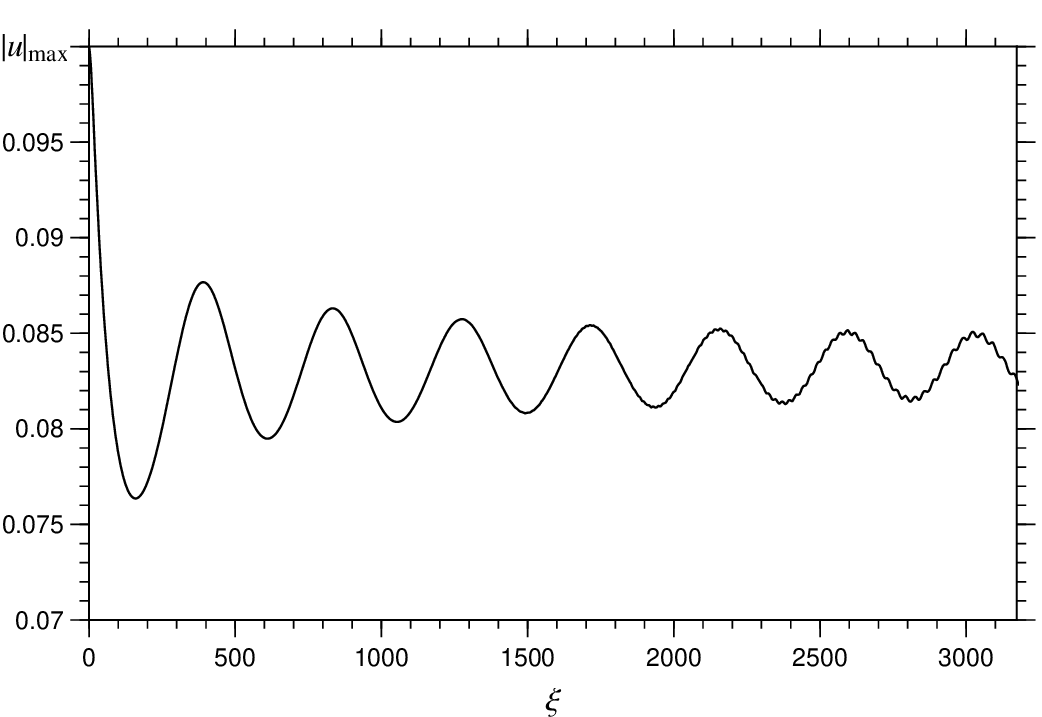}
 \vskip-2mm\caption{\label{HONLSE_kh3_umax}Variations in the amplitude of the quasi-soliton solution with distance at $kh = 3$}
\end{figure}

\begin{figure}[!]
 \vskip1mm
 \includegraphics[width=\column]{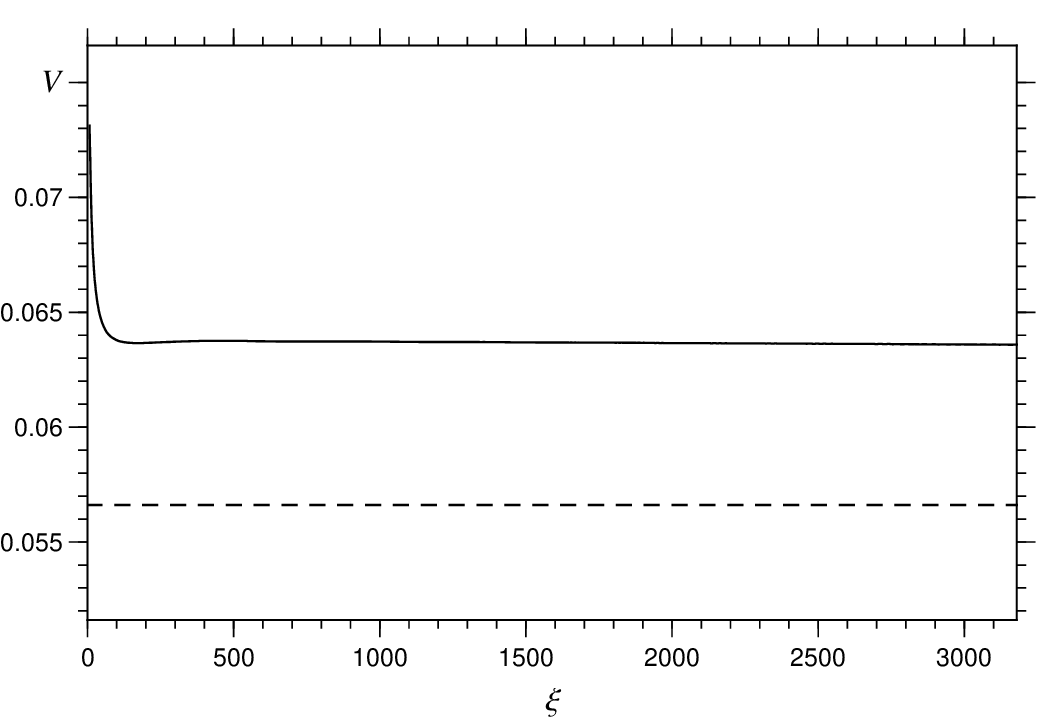}
 \vskip-3mm\caption{\label{HONLSE_kh3_v}Mean wave speed as a function of distance at $kh = 3$: solid curve~--- quasi-soliton, dashed line~--- NLSE soliton ($V\approx $ $\approx0.0566$)}\vspace*{-1mm}
\end{figure}

Figure \ref{fig:HONLSE_kh3} shows the evolution of the same
one-soliton waveform taken as the initial condition in HONLSE
\refe{Eq_u}.\,\,As compared to the NLSE case, the wave amplitude is
smaller, the pulse width is larger, and the wave speed is higher.
The wave amplitude does not remain constant and exhibits slow
oscillations that can be interpreted as the secondary modulation of
the carrier wave.\,\,The amplitude of these oscillations decreases
with time (Fig.\,\,\ref{HONLSE_kh3_umax}).\,\,Such a solution does
not fall under the definition of soliton because it does not
preserve the constant amplitude and shape during the
evolution.\,\,On the other hand, it moves with nearly the constant
speed (Fig.\,\,\ref{HONLSE_kh3_v}) and still possesses the unique
property of solitons to exist over long periods of time without
breaking.\,\,In view of this unique property, we call such solutions
quasi-solitons. The term quasi-soliton was introduced earlier by
Zakharov and Kuznetsov \cite{Zakharov_1998}, but in somewhat
different context; then Karpman et al.\,\,\cite{Karpman_2001} and
Slunyaev \cite{Slunyaev_2009} used it in the same context as in the
present study.


\begin{figure*}[!]
 \vskip1mm
 \includegraphics[width=11.1cm]{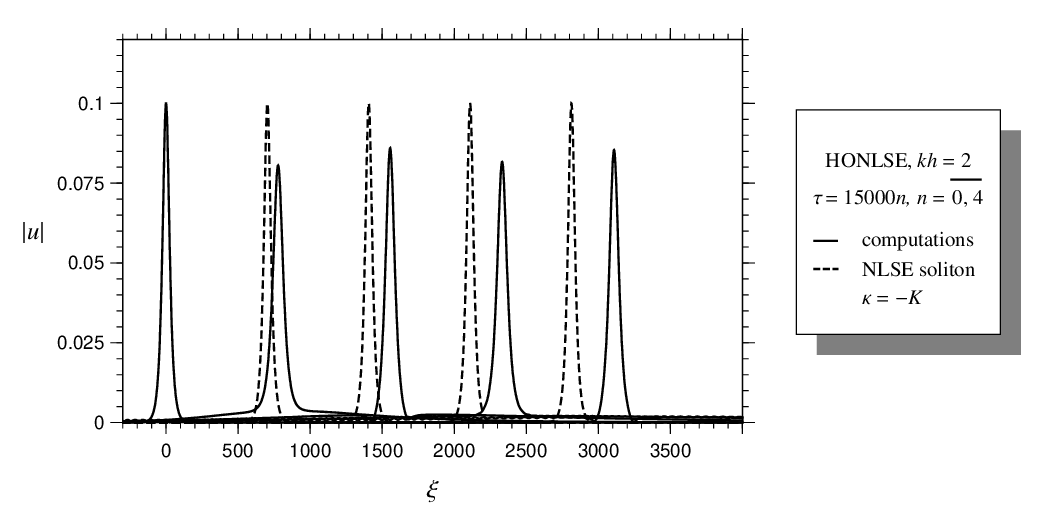}
\vskip-3mm\parbox{11.2cm}{\caption{\label{fig:HONLSE_kh2} Evolution
of one-soliton waveform (\ref{soliton}) taken as the initial
condition in HONLSE \refe{Eq_u} at $kh=2$.\,\,SSF parameters:
$\Delta\tau = 0.5,\; \Delta\xi = 2,\;\xi\in[-4000,\,4000)$.
Accuracy: $S^{(2)}|_{\tau=60000}:\;\mathrm{Er}(I_0) =
 0.028\%$}}\vspace*{-1.5mm}
\end{figure*}

\begin{figure}[!]
 \includegraphics[width=7.8cm]{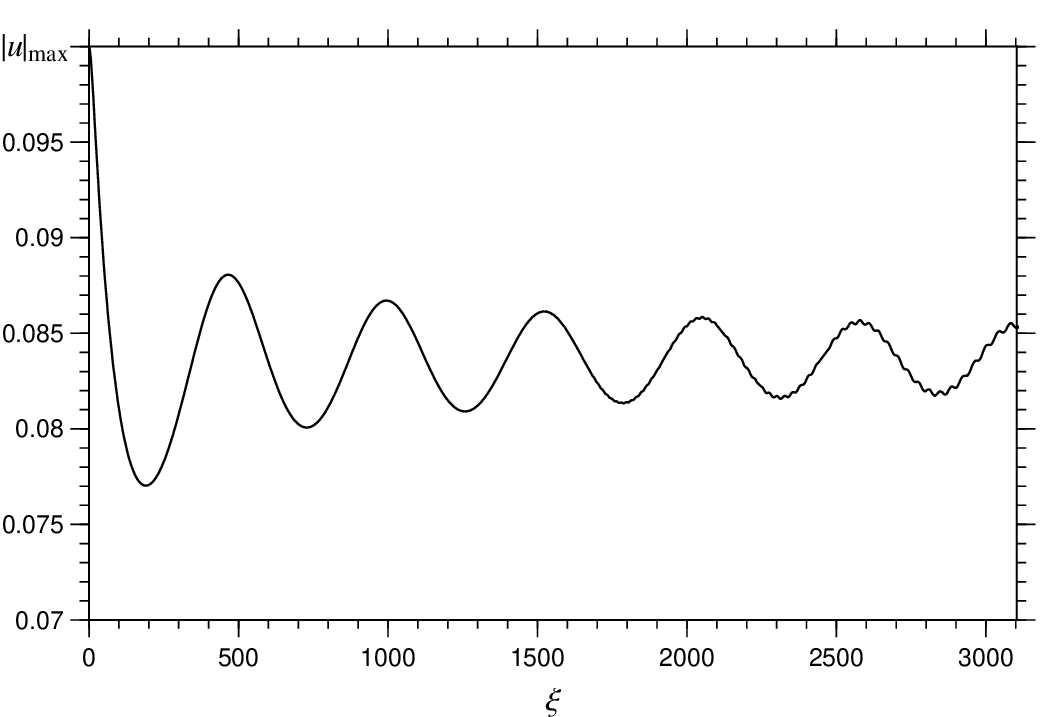}
 \vskip-3mm\caption{\label{HONLSE_kh2_umax}Variations in the amplitude
  of the quasi-soliton solution with distance at $kh = 2$}
\end{figure}

\begin{figure}[h!]
 \includegraphics[width=\column]{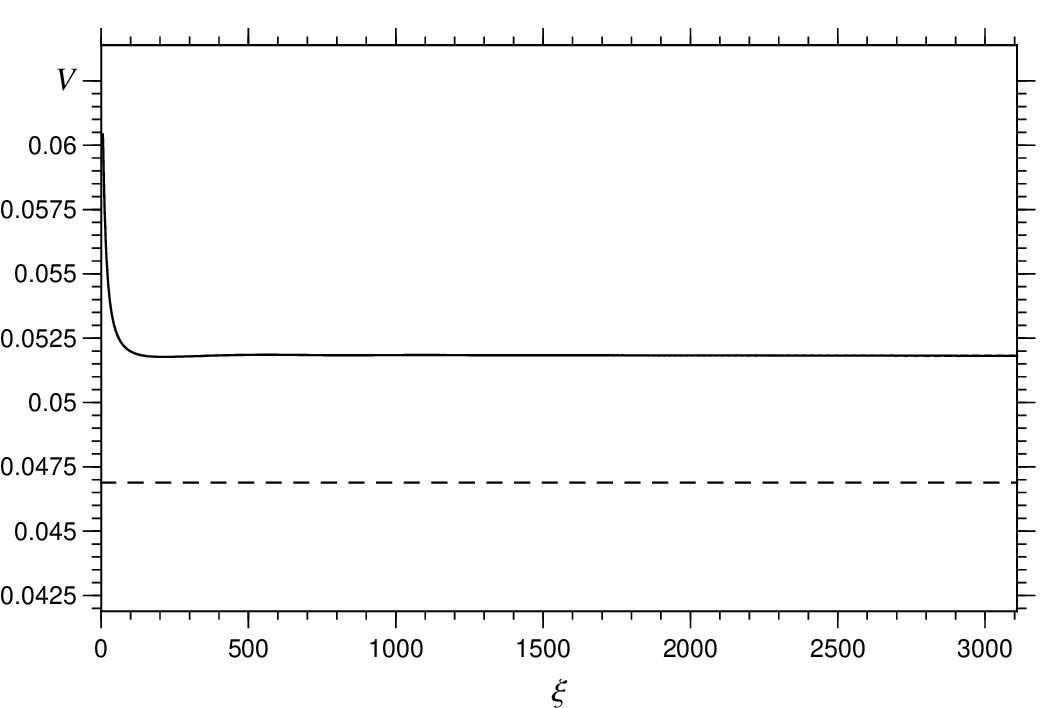}
 \vskip-3mm\caption{\label{HONLSE_kh2_v}Mean wave speed as a
 function of distance at $kh = 2$: solid curve~-- quasi-soliton,
  dashed line~-- NLSE soliton ($V\approx$ $\approx 0.0469$)}%
\end{figure}

Such a behavior of NLSE solitons in the HONLSE case was first
described by Akylas \cite{Akylas_1989} in the context of asymptotic
modeling and numerical simulations of Dysthe's equation in the
infinite-depth limit.\,\,Growth in the soliton speed corresponds to
the well-known carrier frequency downshift observed in deep-water
experiments by Su \cite{Su_1982} and in simulations of Dysthe's
equation by Lo and Mei \cite{Lo_Mei_1985}.\,\,Dysthe
\cite{Dysthe_1979} pointed out that this phenomenon originates due
to the wave-induced mean flow, whose component in the direction of
propagation of the wave causes a local Doppler shift.\,\,Here, we
proved for the first time that this well-known phenomenon can be
observed on finite depth as well.\,\,This result is the main
practical achievement of our
study.\,\,Figures\,\,\ref{fig:HONLSE_kh2}, \ref{HONLSE_kh2_umax},
and \ref{HONLSE_kh2_v} demonstrate that the same quasi-soliton
solution and frequency downshift are observed at a smaller depth,
\mbox{$kh =
2$.}

\begin{figure}[h!]
 \vskip1mm
 \includegraphics[width=\column]{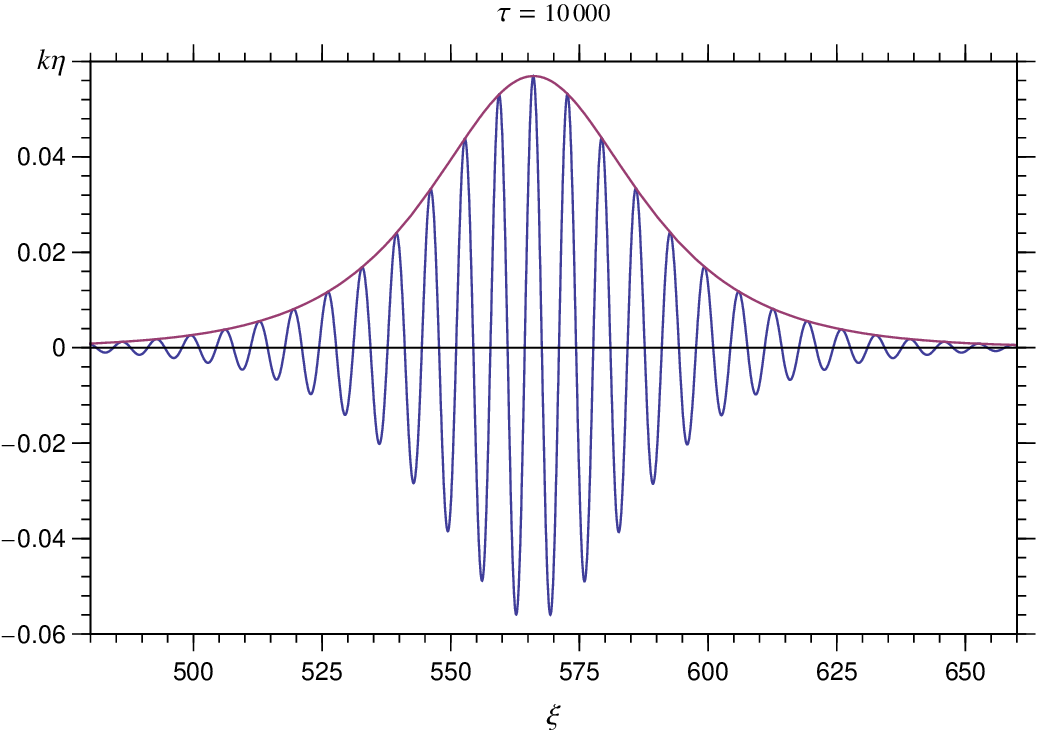}
 \vskip-3mm\caption{\label{HONLSE_kh3_wave}Free surface profile
 with envelope in the form of quasi-soliton at $kh = 3$}\vspace*{-6mm}
\end{figure}
\begin{figure}[!]
 \includegraphics[width=\column]{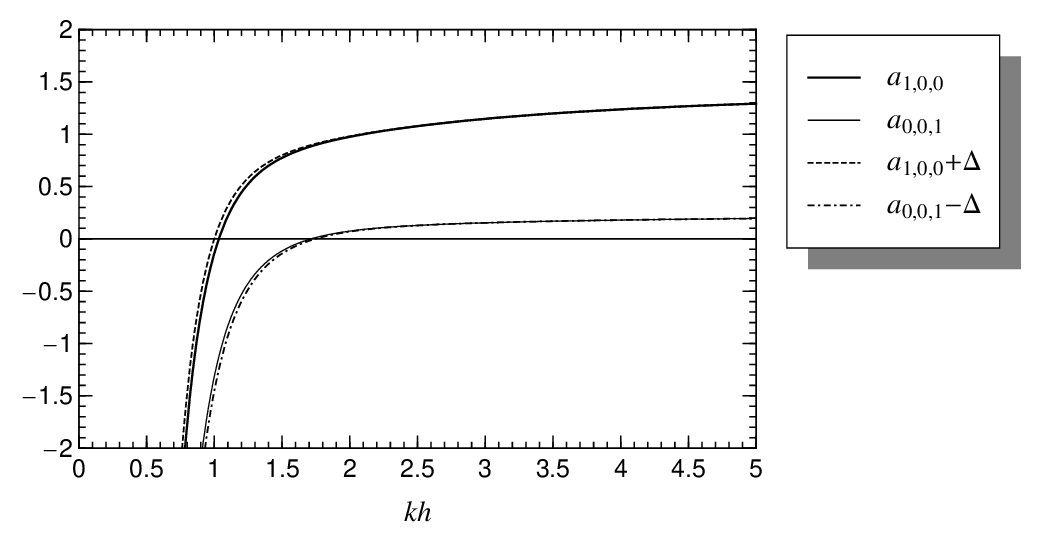}
 \vskip-3mm\caption{\label{fig:delta}Effect of correction
  $\Delta$ on the coefficients $a_{1,\,0,\,0}$ and $a_{0,\,0,\,1}$}
\end{figure}

\begin{figure}[!]
 \vskip3mm
 \includegraphics[width=\column]{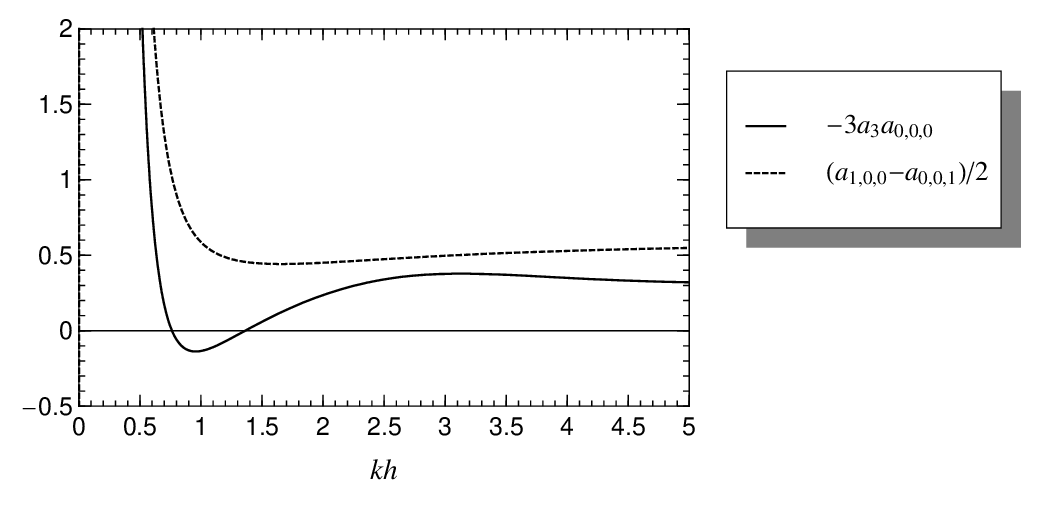}
 \vskip-4mm\caption{\label{fig:an_Satsuma}The left- and
 right-hand-sides
  of Eq.~(\ref{eq:coef_SS}) versus $kh$}\vspace*{1mm}
\end{figure}

Finally, the free surface profile reconstructed with formula
\refe{Eq:zeta} is shown in Fig.~\ref{HONLSE_kh3_wave} for $kh=3$.
The dimensionless maximum free surface elevation is about $0.046$.
The case $kh=3$ corresponds to wavelengths twice as large as depth,
$\lambda\approx 2h$.\,\,The typical depth of the shelf near the
north-west shore of the Black Sea varies from 10 to 100 m.\,\,Hence,
the wavelength corresponding to $kh=3$ falls within the range from
20 to 200~m, which is quite typical of water waves observed on the
Black Sea.\,\,For $h=30$~m, we have $\lambda\approx60$~m and $k =
0.1\;\mathrm{m}^{-1}$.\,\,The corresponding maximum free surface
elevation of the wave shown in Fig.~\ref{HONLSE_kh3_wave} is about
$0.5$~m, and the significant wavetrain width is about 2~km.\,\,Thus,
quasi-soliton solutions obtained in this study can describe swells
propagating on the relatively calm background on seas with
intermediate depths.\,\,The typical trough-to-crest height of such
swells is about 1~m.


\section{Conclusions}\label{sec:concl}

The HONLSE derived earlier by Sedletsky \cite{SedletskyUJP2003} for
the first-harmonic envelope of slowly modulated gravity waves on the
surface of finite-depth irrotational, inviscid, and incompressible
fluid with flat bottom was rewritten in the dimensionless form
suitable for numerical simulations.\,\,One-soliton solutions to NLSE
are transformed into quasi-soliton solutions with slowly varying
amplitude when the HONLSE terms are taken into
consideration.\,\,These quasi-solitons represent the secondary
modulations of gravity waves.\,\,They propagate with nearly constant
speed and possess the unique property of solitons to exist over long
periods of time without breaking.\,\,Their speed was found to be
higher than the speed of the NLSE solitons taken as initial
conditions in computations.\,\,This phenomenon was observed earlier
both in experiment and numerical modeling in the case of deep-water
limit \cite{Su_1982,Akylas_1989}.\,\,It is related to the frequency
downshift originating due to the wave-induced mean flow
\cite{Dysthe_1979,Lo_Mei_1985}.\,\,The quasi-soliton solutions
obtained in this study describe swells propagating on the relatively
calm background on seas with intermediate water depth.

 \vskip3mm

\textit{The authors are grateful to Dr.\,\,S.\,S.~Rozhkov for
initial discussions that motivated us to undertake this study.
D.~Dutykh would like to acknowledge the hospitality of Institut
f\"ur Analysis, Johannes Kepler Universit\"at Linz, where this work
was performed.}


\appendix\footnotesize

\section{On the Correction Introduced by Slunyaev}\label{app:slu}

Slunyaev \cite{Slunyaev_2005} re-derived HONLSE~\refe{eq:Sedletsky}
and introduced a correction,
\[
\Delta
=-\frac{1}{16\sigma^3\nu}\Bigl(\!\bigl(\sigma^2-1\bigr)^4\bigl(3\sigma^2+1\bigr)k^3h^3\,-
\]
\[
 -\,\sigma\bigl(\sigma^2-1\bigr)^2\bigl(5\sigma^4-18\sigma^2-3\bigr)k^2h^2\,+
\]
\begin{equation}
 + \,\sigma^2\bigl(\sigma^2-1\bigr)^2\bigl(\sigma^2-9\bigr)kh +
\sigma^3\bigl(\sigma^2-1\bigr)\bigl(\sigma^2-5\bigr)\!\Bigr),
\end{equation}
to the coefficients $Q_{41}=a_{1,\,0,\,0}$ and
$Q_{42}=a_{0,\,0,\,1}$ derived earlier by Sedletsky
\cite{SedletskyUJP2003}. Actually, this correction was deliberately
ignored by Sedletsky in view of its smallness. Indeed,
Fig.~\ref{fig:delta} proves that $\Delta$ can frankly be ignored at
$kh \gtrsim 1$.

\section{Relationship to the Sasa--Satsuma Equation}\label{app:ss}

Taking into account that $(|u|^{2})_{\xi}=u_{\xi}u^*+ u u^*_{\xi}$,
Eq.~(\ref{Eq_u}) can be rewritten in another form:
\[
 u_{\tau} = -\I a_2 u_{\xi\xi}+ \I a_{0,\,0,\,0}|u|^2 u \,+
\]
\begin{equation}
\label{Eq_u_2} + \left(a_3u_{\xi\xi\xi} -
\widetilde{a}_{1,\,0,\,0}|u|^{2}u_{\xi} -
a_{0,\,0,\,1}u\,(|u|^{2})_{\xi}\right)\!,
\end{equation}
\[
\widetilde{a}_{1,\,0,\,0}=a_{1,\,0,\,0}-a_{0,\,0,\,1}.
\]
When
\begin{equation}\label{eq:coef_SS}
3(-a_3)a_{0,\,0,\,0} = \frac{1}{2}\widetilde{a}_{1,\,0,\,0},
\end{equation}
Eq. (\ref{Eq_u_2}) is reduced to the Sasa--Satsuma equation
\cite{Gilson_2003}, which possesses an infinite number of integrals
of motion and admits some additional exact multi-soliton solutions
in contrast to HONLSE with arbitrary coefficients
\cite{Bandelow_2012}. However, it is clearly shown in
Fig.~\ref{fig:an_Satsuma} that the above relationship among the
parameters is not satisfied for any $kh$. Therefore, the
Sasa--Satsuma equation cannot be obtained from Eq.~(\ref{Eq_u}).

\section{Conservation of the Wave Action}\label{app:cl}

Multiply Eq. (\ref{Eq_u_2}) by $u^*$ and the conjugate equation
by~$u$,
\[
u_{\tau}=-\I a_2 u_{\xi\xi}+ \I a_{0,\,0,\,0}|u|^2 u\,+
\]
\[
 +\, \bigl(a_3 u_{\xi\xi\xi} - \widetilde{a}_{1,\,0,\,0}|u|^{2}u_{\xi}
  - a_{0,\,0,\,1}u\,(|u|^{2})_{\xi}\bigr),
 \; \vert \times u^*,
\]
 \[
u^*_{\tau}=\I a_2 u^*_{\xi\xi}- \I a_{0,\,0,\,0}|u|^2 u^*+
\]
\[
 +\, \bigl(a_3 u^*_{\xi\xi\xi} - \widetilde{a}_{1,\,0,\,0}|u|^{2}u^*_{\xi}
  - a_{0,\,0,\,1}u^*\,(|u|^{2})_{\xi}\bigr),
 \; \vert \times u,
\]
and add these two equations:
\[
\left(u^*u_{\tau}+u u^*_{\tau}\right)=-\I a_2\left(u^*u_{\xi\xi}-u
u^*_{\xi\xi}\right)+
\]
\[
 +\; a_3\left(u^* u_{\xi\xi\xi}+u u^*_{\xi\xi\xi}\right) -
\]
\[
 -\; \widetilde{a}_{1,\,0,\,0}\left(|u|^{2} u^* u_{\xi}+|u|^{2} u u^*_{\xi}\right)
 - 2a_{0,\,0,\,1}|u|^{2}(|u|^{2})_{\xi}.
\]
After some algebraic transformations, we have
\[
\left(|u|^2\right)_{\tau}=-\I a_2\left(\left(u^*u_{\xi}\right)_{\xi}
 -\left(u u^*_{\xi}\right)_{\xi}\right)+
\]
\[
 +\,a_3\left(\left(u^* u_{\xi\xi}\right)_{\xi}-\left(u_{\xi}u^*_{\xi}\right)_{\xi}+
 \left(u u^*_{\xi\xi}\right)_{\xi}\right) -
\]
\[
 -\,\frac{1}{2}\left(\widetilde{a}_{1,\,0,\,0}+2a_{0,\,0,\,1}\right)(|u|^{4})_{\xi}.
\]
In the last term we took into account the following relation
\[
|u|^{2}(|u|^{2})_{\xi} = uu^*(uu^*)_{\xi}=\frac{1}{2}(uu^*uu^*)_{\xi}=\frac{1}{2}(|u|^{4})_{\xi}.
\]
Integrating this equation over $\xi$ from $-\infty$ to $\infty$
yields
\begin{equation}
\int\limits_{-\infty}^{\infty }\left(|u|^{2}\right)_{\tau}\D\xi=0
\;\;\Leftrightarrow\;\; I_0 =
\int\limits_{-\infty}^{\infty}|u|^{2}\D\xi=\mathrm{const},
\end{equation}
where we used the fact that the function $u$ vanishes at $\pm\infty$
along with its derivatives.

%
%
%
%

\section{Split-Step Fourier Technique}\label{app:ssf}
\subsection{Linear equation}

Consider the linear part of HONLSE~\refe{Eq_u}:
\begin{equation}\label{eq:HONLSE_linear}
u_{\tau} = -\I a_2 u_{\xi\xi}+a_3 u_{\xi\xi\xi},\quad u = u(\xi,\,\tau).
\end{equation}
Apply the Fourier transform to the function $u(\xi,\,\tau)$:
\begin{equation}
  \widehat{u}(\kappa,\,\tau) = \frac{1}{2\pi}\int\limits_{-\infty}^{\infty}u(\xi,\,\tau)\exp(-\I \kappa \xi)\D \xi
  \equiv \mathcal{F}_{\kappa}[u(\xi,\,\tau)].
\end{equation}
The inverse Fourier transform is written as
\begin{equation}
  u(\xi,\,\tau) = \int\limits_{-\infty}^{\infty}\widehat{u}(\kappa,\,\tau)\exp(\I \kappa \xi)\D \kappa \equiv \mathcal{F}^{-1}_{\xi}[\widehat{u}(\kappa,\,\tau)].
\end{equation}
The Fourier transforms of the derivatives of function
$u(\xi,\,\tau)$ are expressed as
\begin{equation}
  \widehat{(u_{\xi})}= \I \kappa \widehat{u},\;\; \widehat{(u_{\xi\xi})}= - \kappa^2 \widehat{u},\;...,\;
  \widehat{(u_{n\xi})} = (\I \kappa)^n \widehat{u}.
\end{equation}
Hence, linear equation \refe{eq:HONLSE_linear} takes the following
form in the Fourier space:
\begin{equation}
\widehat{u}_\tau = \left(- \I a_2 (\I \kappa)^2 + a_3 (\I \kappa)^3\right) \widehat{u} ,\quad \widehat{u}(0) \equiv \widehat{u}_0.
\end{equation}
This ordinary differential equation can easily be integrated,
\begin{equation}
  \widehat{u} = \widehat{u}_0 \exp\bigl((\I a_2 \kappa^2 -\I a_3 \kappa^3)\tau\bigr),
\end{equation}
and the following solution for $u(\xi,\,\tau)$ is obtained:
\begin{equation}\label{eq:SSF_linear}
  u = \int\limits_{-\infty}^{\infty}\widehat{u}_0 \exp\bigl((\I a_2 \kappa^2 -\I a_3 \kappa^3)\tau\bigr)\exp(\I \kappa \xi)\D \kappa.
\end{equation}

\subsection{Nonlinear equation}

Nonlinear equation \refe{Eq_u} can be split into the linear and
nonlinear parts:
\[
u_\tau =-\I a_2 u_{\xi\xi} +\I a_{0,\,0,\,0}u|u|^2+
\]
\[
 + \left(a_3 u_{\xi\xi\xi} - a_{1,\,0,\,0}u_{\xi}|u|^{2} - a_{0,\,0,\,1}u^{2}u^{*}_{\xi}\right)
 \equiv \left(\mathcal{L} + \mathcal{N}\right) u,
\]
where
\begin{equation}
\mathcal{L} \equiv -\I a_2 \partial_{\xi\xi} +a_3
\partial_{\xi\xi\xi},
\end{equation}
\begin{equation}
\mathcal{N} \equiv \I a_{0,\,0,\,0}|u|^2 - a_{1,\,0,\,0}u_{\xi}u^{*}
- a_{0,\,0,\,1}u u^{*}_{\xi}
\end{equation}
are the linear and nonlinear operators, respectively. The
semi-discretization in time is performed as follows:
\[
 \frac{u(\xi,\,\tau+\Delta \tau) - u(\xi,\,\tau)}{\Delta \tau}
 \Bigr|_{\Delta \tau \rightarrow 0}= \left(\mathcal{L}
 + \mathcal{N}\right) u(\xi,\,\tau)\Rightarrow
 \]
\[
\Rightarrow u(\xi,\,\tau+\Delta \tau) \approx u(\xi,\,\tau) + \Delta
 \tau (\mathcal{L} + \mathcal{N}) u(\xi,\,\tau)\approx
\]
\[
 \approx \E^{\Delta \tau(\mathcal{L} + \mathcal{N})}u(\xi,\,\tau),
\]
and the second-order Strang formula for noncommuting operators
\cite{Strang_1968} is used:\pagebreak[0]
\[
 \E^{\Delta \tau(\mathcal{L} + \mathcal{N})} \equiv
 S^{(2)}(\Delta\tau)=
\]
\begin{equation}
 = \exp{\Bigl(\!\frac{\Delta \tau}{2}\mathcal{N}\!\Bigr)}
 \exp{\bigl(\Delta \tau\mathcal{L}\bigr)}
 \exp{\Bigl(\!\frac{\Delta \tau}{2}\mathcal{N}\!\Bigr)}\!,\label{eq:NLN}
\end{equation}
\begin{equation}
 =\exp{\Bigl(\!\frac{\Delta \tau}{2}\mathcal{L}\!\Bigr)}
  \exp{\bigl(\Delta \tau\mathcal{N}\bigr)}
  \exp{\Bigl(\!\frac{\Delta \tau}{2}\mathcal{L}\!\Bigr)}\!.\label{eq:LNL}
\end{equation}
In our computations, splitting \refe{eq:LNL} proved to be more
accurate than \refe{eq:NLN}. The linear part\, is\, integrated\,
exactly\, using

\noindent relation~\refe{eq:SSF_linear}
\[
 \E^{\Delta \tau\mathcal{L}} u(\xi,\,\tau)=
\]
\begin{equation}
 =\mathcal{F}_{\xi}^{-1}\bigl[\E^{\Delta \tau \left(-\I a_2 (\I\kappa)^2 +a_3 (\I\kappa)^3 \right)}\mathcal{F}_{\kappa}[u(\xi,\,\tau)]\bigr],
\end{equation}
and the nonlinear part is corrected at each step as follows:
\[
 \E^{\Delta \tau\mathcal{N}} u(\xi,\,\tau)=
 \]
\begin{equation}
= \E^{\Delta \tau \left(\I a_{0,\,0,\,0} |u|^2 -
a_{1,\,0,\,0}u_{\xi}u^{*}
  - a_{0,\,0,\,1}u u^{*}_{\xi}\right)}u(\xi,\,\tau).
\end{equation}
Following Yoshida \cite{Yoshida_1990}, a more accurate fourth-order
splitting can be introduced as well:
\begin{equation}
S^{(4)}(\Delta\tau)=S^{(2)}\left(p_1\Delta\tau\right)S^{(2)}
\left(p_0\Delta\tau\right)S^{(2)}\left(p_1\Delta\tau\right)\!,
\end{equation}
\[
p_0 = -\frac{2^{1/3}}{2-2^{1/3}}\approx -1{.}70,\quad p_1 =
\frac{1}{2-2^{1/3}}\approx 1.35.
\]
For a more detailed description of the SSF technique, the reader can
refer to \cite{Muslu_2005}.

\rezume{І.С.\,Ганджа, Ю.В.\,Седлецький, Д.С.\,Дутих}%
{НЕЛІНІЙНЕ РІВНЯННЯ ШРЕДІНҐЕРА\\ ВИЩОГО ПОРЯДКУ ДЛЯ ОБВІДНОЇ ПОВІЛЬНО\\ МОДУЛЬОВАНИХ ГРАВІТАЦІЙНИХ ХВИЛЬ\\ НА ПОВЕРХНІ РІДИНИ СКІНЧЕННОЇ ГЛИБИНИ\\ ТА ЙОГО КВАЗІСОЛІТОННІ РОЗВ'ЯЗКИ}%
{Розглянуто нелінійне рівняння Шредінґера вищого порядку, виведене
раніше Ю.В.~Седлецьким [УФЖ \textbf{48}(1), 82 (2003)] для обвідної
першої гармоніки повільно модульованих гравітаційних хвиль на
поверхні безвихрової, нев'язкої та нестисливої рідини зі скінченною
глибиною і плоским дном. Це рівняння враховує дисперсію третього
порядку і кубічні нелінійно-дисперсійні доданки. В даній роботі воно
приведено до безрозмірного вигляду, в якому фігурує лише один
безрозмірний параметр $kh$, де $k$~-- хвильове число несучої хвилі,
а $h$~-- незбурена глибина рідини. Показано, що при врахуванні
доданків вищого порядку односолітонні розв'язки класичного
нелінійного рівняння Шредінґера перетворюються в квазісолітонні
розв'язки з повільно змінною амплітудою. Ці квазісолітонні розв'язки
представляють вторинні модуляції гравітаційних хвиль.}

\end{document}